\documentclass[amsmath,aps,prl]{revtex4}
\usepackage{graphicx}
\usepackage{times}
\usepackage{epsfig}

\begin{document}
\title{Influence of external electric fields on multi-photon transitions between the 2s, 2p and 1s levels for hydrogen and antihydrogen atoms and hydrogen-like ions.}

\author{D. Solovyev$^{1}$, L. Labzowsky$^{1,2}$, G. Plunien$^{3}$ and V. Sharipov$^{1}$}

\affiliation{ $^{1}$ V. A. Fock Institute of Physics, St. Petersburg State University, Petrodvorets, Ulianovskaya 1, 198504, St. Petersburg, Russia \\
$^{2}$ Petersburg Nuclear Physics Institute, 188300, Gatchina, St. Petersburg, Russia \\
$^{3}$ Technische Universit\"{a}t Dresden, Mommsenstrasse 13, D-01062, Dresden, Germany.}

\begin{abstract}
One- and two-photon transitions in the hydrogen atom are
analytically evaluated in the absence and in the presence of an
external electric field. The emission probabilities are different
for the hydrogen (H) and anti-hydrogen ($\overline{{\rm H}}$)
atoms due to the existence of contributions, linear in electric
field. The magnitude of these contributions is evaluated within
the nonrelativistic limit. The Coulomb Green function method is
applied. Different nonrelativistic "forms" for the decay
probabilities in combination with different gauge choices are
considered. The three-photon E1E1E1 2p-1s transition probability
is also evaluated and possible applications of the results are
discussed.
\\
PACS number(s): 31.30. Jv, 12.20. Ds, 31.15.-p
\end{abstract}
\maketitle

\section{Introduction}

Recent experimental success in the production of anti-hydrogen
atoms \cite{1,2} opens realistic chances for a comparison of the
atomic spectra for hydrogen (H) and anti-hydrogen ($\overline{{\rm
H}}$) atoms. One of the purposes for this comparison is the search
for the CPT-violating effects \cite{3}. The possibility of the
CPT-tests is connected with the modern extra-accurate resonance
frequency measurements in hydrogen \cite{4,5}. In
\cite{6}-\cite{8} it was shown that a specific difference in the H
and $\overline{{\rm H}}$ atomic spectra arises even in the absence
of the CPT-violation, if an external electric field is present. In
principle, a difference arises also for the frequency measurements
if nonresonant (NR) corrections are taken into account. Actually,
the NR corrections define the natural limit up to which the
frequency measurements have sense: beyond this limit the spectral
line profile cannot be defined by the two parameters: resonance
frequency and width. NR corrections are not just corrections to
the level energy, but rather corrections to the spectral line
profile. These corrections for the atoms placed in an external
electric field contain terms, linear in the electric field.
Therefore NR corrections appear to be different for  H and
$\overline{{\rm H}}$ atoms. However, this difference is very small
and hence the existence of the electric stray field should not
become a serious problem in performing  experiments searching for
CPT-violating effects. Another difference in spectroscopic
properties of H and $\overline{{\rm H}}$ atoms in an external
electric field is the difference in the transition probabilities,
also caused by the terms, linear in electric field. This effect is
much larger than the difference in the NR corrections and was not
yet discussed in detail.

In the present paper we will investigate specially the difference
in the one-photon  2s-1s emission probabilities in H and
$\overline{{\rm H}}$ atoms in an external electric field and will
find the optimal conditions for observing this effect. This
investigation should confirm our understanding of fundamental
symmetries in physics. Thus, going over from H to $\overline{{\rm
H}}$ and changing simultaneously the sign of the electric field
should not change the atomic spectra provided that CPT symmetry is
conserved.

The one-photon decay of the 2s state of the H atom in an external
electric field was utilized in the experiments \cite{4,5} for the
registration of the 2s-1s two-photon resonance absorption. The
difference of the one-photon 2s-1s decay rate for H and
$\overline{{\rm H}}$ atoms in an external electric field can be
observable in the similar experiments with $\overline{{\rm H}}$
atoms.

The probabilities for the spontaneous two-photon decay in hydrogen
atoms and hydrogen-like ions are under investigation since the
theoretical formalism has been introduced by G\"{o}ppert-Mayer
\cite{Goppert} and the first evaluation for the two-photon E1E1
transition ${\rm 2s}\rightarrow 2\gamma({\rm E1})+{\rm 1s}$ has
been presented by Breit and Teller \cite{Breit}. A highly accurate
calculation of the E1E1 - transition probability has been
performed by Klarsfeld \cite{Klarsfeld}. Recently Jentschura
\cite{Jentschura} performed a complete evaluation of the radiative
corrections and presented more accurate value of the E1E1
two-photon decay probability. The double- and triple-photon decays
of metastable ${}^3P_0$ atomic state were considered in
\cite{Schmieder}. The present paper is devoted also to evaluation
of the probabilities for two-photon decays ${\rm 2p}\rightarrow
\gamma({\rm E1})+ \gamma({\rm M1})+{\rm 1s}$ and ${\rm
2p}\rightarrow \gamma ({\rm E1})+\gamma({\rm E2})+{\rm 1s}$.
Evaluations of these two transitions have been first accomplished
in \cite{LabShon,LabShonSol} for hydrogen-like systems with
nuclear charge numbers $Z$ within the range $1\leq Z\leq 100$ by
pure numerical methods. Here we present analytic calculations in
the nonrelativistic limit and compare them with corresponding
numerical results. For performing the summations over intermediate
states (i.e. over the complete set of solutions of the
Schr\"{o}dinger equation describing electrons in the Coulomb field
of the nucleus) we employ the Coulomb Green function
\cite{Hostler}. The Green function method has been first applied
for deriving the general expression for the two-photon decay
probability in  H  atom and  H-like ions in \cite{ZR,RZM}. An
alternative approach applicable for arbitrary states based on
Schwinger's analytical representation of the Coulomb Green
function has been presented in \cite{Kelsey,Gasean}.

As it has been indicated in \cite{Drak}, the nonrelativistic
behavior of E1M1 transitions as a function of $Z$ with the neglect
of the interelectron interaction should be $W^{\rm
E1M1}=(8/9\pi)(\alpha Z)^{12}/100$ for the helium-like systems.
This very small value arises due to the cancellation of
contributions of the leading terms 2p$_{1/2}$ and 2p$_{3/2}$ in
the summation over intermediate np-states. However, we should note
that this result yields only a minor contribution for small
nuclear charges $Z$ when it is evaluated within the ``velocity''
gauge \cite{Drak}. In this case a major contribution arises from
the negative-energy intermediate states and scales like $(\alpha
Z)^8$ in atomic units \cite{LabShon}, \cite{LabShonSol}.

The two-photon ${\rm 2p}_{1/2}\rightarrow \gamma ({\rm
E1})+\gamma({\rm E2}) + {\rm 1s}$ transition rate for hydrogen and
hydrogen-like ions is proportional to $(\alpha Z)^8$ in atomic
units. Thus the two-photon transitions represent higher-order
corrections to the life time of the 2p$_{1/2}$-level when compared
to the lowest order $(\alpha Z)^3 \ln (\alpha Z)$ (in relativistic
unit) radiative corrections derived in references \cite{Karsh},
\cite{Karshen}. A direct observation of the influence of the two
photon ${\rm 2p}_{1/2}\rightarrow \gamma ({\rm E1})+\gamma({\rm
E2}) + {\rm 1s}$ transitions in the H atom does not look feasible
due to the huge background arising from the one-photon transition
${\rm 2p}_{1/2}\rightarrow \gamma({\rm E1})+{\rm 1s}$. However,
two-photon decays of the 2p$_{1/2}$-level could be observed in
coincidence experiments.

In this paper we reevaluate the two-photon decay rates of the
2p-state in hydrogen, deriving the E1E2 and E1M1 contributions to
the two-photon emission processes by means of analytical methods.
These calculations are performed within different gauges and
employing different forms for the expression of the transition
probability (see \cite{LabSol}, \cite{NikRudz}).

We also perform the calculation of two-photon transition
probabilities from the 2s and 2p$_{1/2}$ hydrogenic levels in the
presence of an external electric field. The mixing of the 2s and
2p$_{1/2}$ levels results in additional terms, linear in electric
field. The evaluation is performed within the nonrelativistic
limit by the Coulomb Green function method \cite{RZM} in the
"length" gauge. These terms linear in the electric field lead to
the difference of the corresponding probabilities, see
\cite{Azimov}-\cite{Hillery}, for the hydrogen and anti-hydrogen
atoms. The magnitude of these contributions is evaluated and the
possibility of the observation of this effect is discussed. It is
important to stress, that these linear field corrections even in a
very small stray electric field can be larger than the radiative
correction to the decay of 2s level, evaluated in
\cite{Jentschura}.

Finally, we evaluate a three-photon emission probability for the
hydrogen atom. This question was never debated in literature.

The recent success in observation of the cosmic microwave
background temperature and polarization anisotropy draw attention
to the details of the cosmological hydrogen recombination history.
This, in turn, required the accurate knowledge of the two-photon
decay processes in hydrogen (see \cite{Chluba,J.Chluba} for
details and references). The main interest concentrates around the
2s-1s E1E1  two-photon transition and the two-photon decays for
the ns, nd excited states.

The smallness of the numerical coefficients in E1E2 and E1M1
two-photon expressions excludes any significant role of these
transitions for astrophysics. The same can be stated for the
3-photon 3E1 2p$_{1/2}-$1s decay. However the values for the
2p$_{1/2}-$1s E1M1 and E1E2 transition rates appeared to be
important for the determination of the theoretical accuracy limits
\cite{LabShedr,LabShedr1} for frequency measurements by means of
optical resonance experiments with hydrogen \cite{4,5}.

The following notations will be used throughout this paper. The
2p$_{1/2}$ level is labelled as 2p because the 2p$_{3/2}$ level
will not be considered. Vectors with 3 components are in bold.
Angular part of any vector ${\bf r}$ is defined by the unit vector
${\bf n}_{\bf r}$ directed the same as ${\bf r}$. The
Clebsh-Gordon coefficients $C_{\ell_1 m_1\,\ell_2 m_2}^{jm}$ are
introduced according to \cite{Varsh}. Other notations are the
following:
\begin{eqnarray}
\Delta = \sqrt{\Delta E^{2}_{\rm L}+\frac{1}{4}\Gamma^{2}_{\rm 2p}},\\
w^{(1\gamma)} = \sqrt{W^{(1\gamma)}_{\rm 2s}/W^{(1\gamma)}_{\rm 2p}},\\
w^{(2\gamma)} = \sqrt{W^{(2\gamma)}_{\rm 2s}/W^{(2\gamma)}_{\rm 2p}}.
\end{eqnarray}
Here $\Delta E_{\rm L}$ is the Lamb shift, $\Gamma^{}_{\rm 2p}$ is
the width of the 2p state, $W^{(1\gamma),(2\gamma)}_{\rm 2s}$ and
$W^{(1\gamma),(2\gamma)}_{\rm 2p}$ are the one-photon and
two-photon decay probabilities for the corresponding 2s and 2p
levels. The value $W^{(2\gamma)}_{\rm 2p}=W^{\rm (E1E2)}_{\rm
2p}+W^{\rm (E1M1)}_{\rm 2p}$ includes the probabilities for ${\rm
2p}\rightarrow \gamma ({\rm E1})+\gamma({\rm E2}) + {\rm 1s}$ and
${\rm 2p}\rightarrow \gamma ({\rm E1})+\gamma({\rm M1}) + {\rm
1s}$ decays.

\section{2s decay rate for hydrogen and anti-hydrogen atoms in external electric fields}

In the absence of an electric field, the M1 transition ${\rm
2s}\rightarrow {\rm 1s}+\gamma$ - strongly forbidden in the
nonrelativistic limit - was first evaluated by Breit and Teller
\cite{Breit}. The value, obtained in \cite{Breit} was later
improved by Drake \cite{Drake} (see also \cite{Sucher}). In
\cite{Drake,Sucher} relativistic corrections to the
Schr\"{o}dinger wave functions and to the M1 photon emission
operator were taken into account. Neglecting these corrections the
${\rm 2s}\rightarrow {\rm 1s}+\gamma$ transition probability turns
to be zero due to the orthogonality of the radial wave functions.
Accurate fully relativistic calculations for the ${\rm
2s}\rightarrow {\rm 1s}+\gamma$ transition for the hydrogen-like
ions with arbitrary nuclear charge Z values were performed by
Johnson \cite{Johnson}.

Within QED theory the S-matrix element for the one-photon emission
process reads \cite{Akhiezer}
\begin{equation}\label{1.S-matrix}
S_{A'A}=\frac{e\sqrt{4\pi}}{\sqrt{2\omega}}\int
dx\,\bar{\psi}_{A'}(x)e^{*}_{\mu}\gamma^{\mu}e^{-i({\bf k}{\bf
r}-\omega t)}\psi_A(x),
\end{equation}
where ${\bf k}$ is the wave vector of the photon, $\omega=|{\bf
k}|$ is the photon frequency, $x=({\bf r},t)$ is the 4-vector of
space-time coordinates, $e_{\mu}$ is the 4-vector of the photon
polarization, $\psi_A(x)$ is the Dirac wave function for the bound
electron in an atom, $\gamma_{\mu}$ are the Dirac matrices and
$A'$, $A$ correspond to the relevant quantum numbers
characterizing the final and initial states of the electron in an
atom. In Eq.~(\ref{1.S-matrix}) $e$ is the electron charge and the
relativistic units $\hbar=c=1$ are used throughout this section.

We integrate over the time variable $t$ in Eq.~(\ref{1.S-matrix})
and employ the relation
\begin{equation}\label{SAA}
S_{A'A}=-2\pi i\delta(E_{A'}-E_A-\omega)U_{A'A}
\end{equation}
for the transition amplitude $U_{A'A}$. In what follows, we define
also the amplitude $U_{A'A}({\bf k},{\bf e})$ as
\begin{equation}\label{1.3}
U_{A'A}({\bf k},{\bf e})=\frac{\sqrt{2\pi\omega}}{e}\,U_{A'A} =
\langle A'|{\bf e}{\bf \alpha}e^{-i{\bf k}{\bf r}}|A\rangle.
\end{equation}
Here ${\bf e}$ is the transverse polarization vector and
${\bf \alpha}$ are the Dirac matrices.

Transition probability integrated over the photon emission
directions and summed over the polarizations is
\begin{equation}\label{1.first}
W_{A'A}^{(1\gamma)}=\frac{e^2\omega_{AA'}}{2\pi}\sum\limits_{\bf
e}\int d{\bf n}_{\bf k}\,\left|U_{A'A}({\bf k},{\bf e})\right|^2,
\end{equation}
$\omega_{AA'}=E_{A}-E_{A'}$, $E_{A'}$, $E_A$ are the Dirac
eigenvalues for the atomic electron, $\langle A'|...|A\rangle $ is
the matrix element with the Dirac 4-component wave functions.
Summation over ${\bf e}$ implies the transversality condition
${\bf e}{\bf k}=0$.

Within the Pauli approximation Eq.~(\ref{1.first}) reduces to
\cite{Sucher}:
\begin{equation}\label{1.first_a}
W_{A'A}^{(1\gamma)}=\frac{e^2\omega_{AA'}}{2\pi}\sum\limits_{\bf
e}\int d{\bf n}_{\bf k}\, \left|U^{\rm P}_{A'A}({\bf k},{\bf
e})\right|^2
\end{equation}
with  the transition amplitude in the Pauli approximation defined
via the matrix element of the corresponding emission operator
\begin{eqnarray}\label{1.6}
U_{A'A}^{\rm P}({\bf k},{\bf e})=\left(({\bf e}\hat{\bf p}+i{\bf
e}[{\bf k}\times\hat{\bf s}])e^{-i{\bf k}{\bf r}}\right)_{A'A}.
\end{eqnarray}
The relation Eq.~(\ref{1.6}) involves the nonrelativistic electron
momentum operator $\hat{\bf p}=-i\nabla$, the electron spin
operator $\hat{\bf s}=\frac{1}{2}{\bf \sigma}$ (${\bf \sigma}$ are
the Pauli matrices) and $(...)_{A'A} \equiv (A'|...|A)$ denotes
the matrix element with the Schr\"{o}dinger wave functions. The
first term of the integrand in Eq.~(\ref{1.first_a}) describes the
E1 photon emission, that in case of $A'={\rm 1s}$, $A={\rm 2s}$ is
forbidden by parity; The second term corresponds to the M1 photon
emission. Due to the orthogonality of the radial wave functions
$\psi_{\rm 2s}$ and $\psi_{\rm 1s}$ the M1 transition probability
as described by Eq.~(\ref{1.first_a}) is nonzero only due to the
factor $e^{-i{\bf k}{\bf r}}$ and due to the relativistic
corrections to the Schr\"{o}dinger wave functions. For the atomic
electron $r\simeq a_0=1/m\alpha Z$ ($a_0$ is the Bohr's radius,
$m$ is the electron mass, $\alpha\approx 1/137$ is the fine
structure constant), $\omega\approx m(\alpha Z)^2$ and ${\bf
k}{\bf r}\approx\alpha$. Thus, one can restrict the consideration
to the lowest nonvanishing term of the multipole expansion of
$e^{-i{\bf k}{\bf r}}$, which turns to be of the same order of
magnitude as the relativistic corrections to the wave functions.

The zero-order wave functions which should be employed for the
evaluation of the matrix elements in Eq.~(\ref{1.first_a}) look
like
\begin{equation}
\psi_A({\bf r})\equiv\psi_{njlm}({\bf r})=\sum\limits_{m_lm_s}\,
C_{\ell m_\ell\,\frac{1}{2} m_s}^{jm}\, R_{nl}(r)Y_{lm_l}({\bf
n}_{\bf r})\chi_{m_s},
\end{equation}
where the standard set of one-electron quantum numbers is
introduced: Principal quantum number $n$, total electron angular
momentum $j$ and its projection $m$, orbital electron angular
momentum $l$ and its projection $m_l$, spin projection $m_s$. The
function $\chi_{m_s}$ is the Pauli spinor. For the 2s and 1s
electron states $l=0$ and the transition amplitude in
Eq.~(\ref{1.first_a}) reduces to \cite{Sucher}
\begin{eqnarray}\label{1.5}
U^{\rm P}_{{\rm 2s}m_s\,{\rm 1s}m'_s}=-ik^2\langle m_s|{\bf e}[{\bf k}\times{\bf \sigma}]|m'_s\rangle R_{{\rm 2s}\,{\rm 1s}},\\
R_{{\rm 2s}\, {\rm 1s}}=\int\limits_0^{\infty}R_{\rm 2s}(r)R_{\rm
1s}(r)r^4\,dr,
\end{eqnarray}
where $|m_s\rangle$ are the spin wave functions, $R_{\rm 2s}(r)$,
$R_{\rm 1s}(r)$ are the nonrelativistic radial wave functions.
Inserting Eq.~(\ref{1.5}) into Eq.~(\ref{1.first_a}), we have to
sum over $m'_s$ and average over $m_s$. For summation over ${\bf
e}$ the standard formula
\begin{equation}\label{1.Fik}
\sum\limits_{\bf e}e_ie_k=\delta_{ik}-n_i n_k
\end{equation}
is used, where $n_i$ and $n_k$ are the Cartesian components of the
vector ${\bf n}_{\bf k}$. After integration over the photon
emission directions ${\bf n}_{\bf k}$ one finds
\begin{equation}\label{1.prob_a}
W_{{\rm 2s}\,{\rm 1s}}^{(1\gamma)}=\frac{3^5}{2^{21}}\alpha
m^5(\alpha Z)^{14}R^2_{{\rm 2s}\, {\rm 1s}},
\end{equation}
respectively. The evaluation of the radial integral in
Eq.~(\ref{1.prob_a}) results
\begin{equation}
W_{{\rm 2s}\,{\rm 1s}}^{(1\gamma)}=\frac{1}{972}m\alpha(\alpha
Z)^{10}=2.8\cdot 10^{-6}\,\, {\rm s}^{-1}\, .
\end{equation}
The decay rate of the 2s state of a hydrogen atom in the presence
of an external electric field was studied in
\cite{Azimov}-\cite{Hillery}. The external electric field mixes
the states 2s and 2p. A 100\% mixing occurs in a field with the
strength $D_{\rm c}=475$ V/cm \cite{Bethe}. In what follows we
will consider weaker fields $D<D_{\rm c}$ such that admixtures of
all other states, apart from 2p can be neglected. The ground state
1s will be assumed unaffected by the field. We will denote the
mixed states 2s and 2p as $\overline{\rm 2s}$ and $\overline{\rm
2p}$, respectively. For the wave function of the $\overline{\rm
2s}$ state we can write \cite{Mohr}
\begin{equation}\label{1.mixing}
|\overline{\rm 2s}m_s\rangle =|{\rm 2s} m_s\rangle
+\eta\sum\limits_{m''_s} \langle  {\rm 2p}m''_s|e{\bf D}{\bf
r}|{\rm 2s}m_s\rangle |{\rm 2p}m''_s\rangle,
\end{equation}
where $\eta=(\Delta E_{\rm L}+i\Gamma_{\rm 2p}/2)^{-1}$, ${\bf D}$
is the electric field vector.

The transition  amplitude Eq.~(\ref{1.5}) in an external electric
field looks like
\begin{equation}\label{1.8}
 U^{\rm P}_{\overline{\rm 2s}m_s\,{\rm 1s}m'_s}({\bf k},{\bf e})=U^{\rm P}_{{\rm 2s}m_s\,{\rm 1s}m'_s}({\bf k},{\bf
 e})+\eta\sum\limits_{m''_s} \langle {\rm 2s}m_s|e{\bf D}{\bf r}|{\rm 2p}m''_s\rangle U^{\rm P}_{{\rm 2p}m''_s\,{\rm 1s}m'_s}({\bf k},{\bf e}),
\end{equation}
where
\begin{equation}\label{1.9}
U^{\rm P}_{{\rm 2p}m''_s\,{\rm 1s}m'_s}({\bf k},{\bf
e})=im\omega_{{\rm 2s}\,{\rm 1s}} \langle {\rm 2p}m''_s|{\bf
e}{\bf r}|{\rm 1s}m'_s\rangle.
\end{equation}
Direct evaluation of the integral in Eq.~(\ref{1.9}) results in
\begin{equation}
U^{\rm P}_{{\rm 2p}m''_s\,{\rm 1s}m'_s}({\bf k},{\bf
e})=3im\omega_{{\rm 2s}\,{\rm 1s}}\sum\limits_q(-1)^qe_q
C_{1\bar{q}\,\frac{1}{2}m'_s}^{\frac{1}{2}m''_s},
\end{equation}
where $e_q$ are the spherical components of the vector ${\bf e}$
and $\bar{q}=-q$. Similarly,
\begin{equation}\label{DmElem}
\langle {\rm 2s}m_s|e{\bf D}{\bf r}|{\rm 2p}m''_s\rangle
=3e\sum\limits_{q'}(-1)^{q'}D_q
C_{1\bar{q}'\,\frac{1}{2}m'_s}^{\frac{1}{2}m''_s}.
\end{equation}
Further evaluation requires the insertion of the amplitude
Eq.~(\ref{1.5}) in Eq.~(\ref{1.first}) and summation over ${\bf
e}$. For this purpose the formula Eq.~(\ref{1.Fik}) should be used
and then the representation of the scalar and vector products in
spherical components should be employed. The final result is (here
we do not integrate over the directions ${\bf n}_{\bf k}$):
\begin{equation}\label{1.dW}
dW_{\overline{\rm 2s}\,{\rm 1s}}^{(1\gamma)}({\bf n}_{\bf k})=
\frac{3}{8\pi}W_{{\rm 2s}\, {\rm 1s}}^{(1\gamma)}\left[1+e{\bf
D}{\bf n}_{\bf k}\frac{\Gamma_{\rm
2p}}{w^{(1\gamma)}\Delta^{2}}+\frac{e^2D^2}{(
w^{(1\gamma)}\Delta)^2}\right]d{\bf n}_{\bf k}\, .
\end{equation}
Although formula Eq.~(\ref{1.dW}) was obtained earlier in
\cite{Azimov} and \cite{Mohr}, \cite{Hillery}. We present here our
way for its derivation.

The formal T-noninvariance of the factor ${\bf D}{\bf n}_{\bf k}$
in Eq.~(\ref{1.dW}) (${\bf n}_{\bf k}$ and ${\bf D}$ are T-odd and
T-even vectors, respectively) is compensated by the dependence on
$\Gamma_{\rm 2p}$; This is the imitation of T-noninvariance  in
unstable systems, as predicted by Zeldovich \cite{Zeldovich}.
Rewriting Eq.~(\ref{1.dW}) into the form \cite{Azimov}
\begin{equation}\label{1.19}
dW_{\overline{\rm 2s}\,{\rm
1s}}^{(1\gamma)}=W_{0}\left[1\mp\beta(D){\bf n}_{\bf D}{\bf
n}_{\bf k}\right]d{\bf n}_{\bf k},
\end{equation}
where
\begin{equation}
W_0=\frac{3}{8\pi}W_{{\rm 2s}\, {\rm
1s}}^{(1\gamma)}\left(1+\frac{e^2D^2}{(w^{(1\gamma)}\Delta)^2}\right),
\end{equation}
\begin{equation}
\beta(D)=\frac{|e|D\Gamma_{\rm 2p}
w^{(1\gamma)}}{(w^{(1\gamma)}\Delta)^2+e^2D^2},
\end{equation}
we find the maximum value $\beta(D)$ at \cite{Azimov}
\begin{equation}\label{1.max}
D_{\rm max}=\frac{w^{(1\gamma)}\Delta}{|e|}\approx 0.3\cdot
10^{-4}\,{\rm V/cm}.
\end{equation}
The value $\beta_{\rm max}=\beta(D_{\rm max})$ is equal to
\begin{equation}
\beta_{\rm max}=\frac{\Gamma_{\rm
2p}}{2\Delta}\approx\frac{1}{20}.
\end{equation}
The $(-)$ and $(+)$ signs in Eq.~(\ref{1.19}) correspond to the H
and $\overline{{\rm H}}$ atoms, respectively.

The relative difference for the decay rates in H and $\overline{{\rm H}}$ atoms at the maximum value $D_{\rm max}$ equals to:
\begin{equation}
\frac{dW_{\overline{\rm 2s}\, {\rm 1s}}^{(1\gamma)}({\rm
H})}{dW^{(1\gamma)}_{{\rm 2s}\, {\rm 1s}}}-\frac{dW_{\overline{\rm
2s}\, {\rm 1s}}^{(1\gamma)}(\overline{{\rm
H}})}{dW^{(1\gamma)}_{{\rm 2s}\,{\rm 1s}}}=\frac{W_0(D_{\rm
max})2\beta(D_{\rm max}){\bf n}_{\bf D}{\bf n}_{\bf
k}}{\frac{3}{8\pi}W^{(1\gamma)}_{{\rm 2s}\, {\rm 1s}}}\approx
\frac{1}{5}{\bf n}_{\bf D}{\bf n}_{\bf k}.
\end{equation}
In the presence of such a very weak field given by
Eq.~(\ref{1.max}), this difference is close to about $20$\% and
probably can be observed in experiments of the type reported in
\cite{4,5}.

We would also note that if one integrates in Eq.~(\ref{1.19}) over
photon emission directions, the term linear with respect to the
field vanishes. But a quadratic term included in $W_0$ exists and
represents a correction to the $W^{(1\gamma)}_{{\rm 2s}\, {\rm
1s}}$ transition probability, i.e. to the lifetime of the 2s
level. This correction term reaches the magnitude of the radiative
correction obtained in \cite{Jentschura} in the field of the
strength
\begin{equation}\label{1.corr}
D_{\rm
r}=\frac{1}{|e|}\sqrt{\frac{8\pi}{3}\Delta\frac{\delta\Gamma_{\rm
2s}}{W_{{\rm 2p}\, {\rm
1s}}^{(1\gamma)}}}\approx\sqrt{\frac{8\pi}{3}\frac{\delta\Gamma_{\rm
2s}}{\Gamma_{\rm 2s}}}D_{\rm max},
\end{equation}
where the correction $\delta\Gamma_{\rm 2s}/\Gamma_{\rm 2s}$ was
derived by Jentschura (see Eq.~(36) in \cite{Jentschura}). Though
the correction in \cite{Jentschura} is obtained for the process of
the two-photon decay of 2s level, it also represents a correction
to the lifetime of the 2s level. For the hydrogen atom $D_{\rm
r}\sim 1.4\cdot 10^{-7}$ V/cm. Since the experiments deal with
differential cross sections, we compare also the linear term with
the radiative correction. The linear term $\beta(D)$ reaches the
magnitude of $\delta\Gamma_{\rm 2s}/\Gamma_{\rm 2s}$ at the field
strength $7.5\cdot 10^{-11}$ V/cm. It should be difficult to
eliminate spurious fields of such magnitude in real experiments
and, therefore, the comparison of the theoretical results in
\cite{Jentschura} with experimental ones requires some caution.

\section{Transition probabilities in different forms and gauges}

In this section different gauges in combination with different
"forms" for the one-photon transition probability are described.
Atomic units $\hbar=e=m=1$ will be used throughout this section.

The transition probability for the emission of a photon with
definite angular momentum and parity can be described in the first
order of QED perturbation theory within arbitrary gauge as
\begin{eqnarray}\label{2.1}
W_{A\rightarrow A'}(\omega )&=&\sum\limits_{kq}\left[\left|\langle
A'|\left({\bf \alpha}{}_{{\rm e}}{\bf A}_{\omega
kq}({\bf r})\right)+\Phi_{\omega kq}\left({\bf r}\right)|A\rangle \right|^2\right.\nonumber\\
&& \left. +\left|\langle A'|{\bf \alpha}{}_{{\rm m}}{\bf
A}_{\omega kq }\left({\bf r}\right)|A\rangle \right|^2\right],
\end{eqnarray}
where ${}_{{\rm e}}{\bf A}_{\omega kq}$ and ${}_{{\rm m}}{\bf
A}_{\omega kq}$ denote the electric and magnetic vector potentials
and $\Phi_{\omega kq}$ corresponds to the scalar potential;
$\omega$ is the photon frequency, $k$, $q$ are the total angular
momentum of the emitted photon and its projection. The bra-kets
$|A\rangle $ and $\langle A'|$ are stationary Dirac states (wave
functions) with energies $E_A$ and $E_{A'}$  and  ${\bf \alpha}$
are the Dirac matrices. In the momentum representation these
potentials take the form
\begin{eqnarray}\label{2.2}
{}_{{\rm e}}{\bf A}_{\omega kq}({\bf k})&=&\frac{4\pi^2c^{3/2}}{\omega^{3/2}}\delta\left(k-\frac{\omega}{c}\right)\left({}_{{\rm e}}{\bf Y}_{kq}({\bf n}_{\bf k})+K{\bf n}_{\bf k}Y_{kq}({\bf n}_{\bf k})\right)\, , \\
{}_{{\rm m}}{\bf A}_{\omega kq}\left({\bf k}\right)&=&\frac{4\pi^2c^{3/2}}{\omega^{3/2}}\delta\left(k-\frac{\omega}{c}\right){}_{{\rm m}}{\bf Y}_{kq}({\bf n}_{\bf k})\, , \\
\Phi_{\omega kq}\left(k\right)&=&\frac{4\pi^2c^{3/2}}{\omega^{3/2}}\delta\left(k-\frac{\omega}{c}\right)KY_{kq}({\bf n}_{\bf k})\, .
\end{eqnarray}
Here ${\bf k}$ denotes the variable in the momentum
representation. Functions ${}_{{\rm e}}{\bf Y}_{kq}$ and ${}_{{\rm
m}}{\bf Y}_{kq}$ are the vector spherical harmonics of electric
and magnetic type, respectively, $Y_{kq}$ is the ordinary
spherical harmonic, $c$ is the speed of light and $K$ denotes a
gauge-dependent constant.

The spherical components of the transversal electric ${}_{{\rm
e}}A^{(1\lambda)}$ and the longitudinal ${}_{{\rm
l}}A^{(1\lambda)}$ parts of the electromagnetic vector potential
(the superscript $(1\lambda)$ defines the rank of a spherical
tensor and labels the components) are
\begin{eqnarray}\label{2.3.1}
{}_{{\rm e}}A^{(1\lambda)}_{\omega kq}&=&\sqrt{\frac{\omega}{\pi
c(2k+1)}}\left[\sqrt{k(2k+3)}\left(
\begin{array}{ccc}
1&k+1&k\\
\lambda&q-\lambda&q
\end{array}
\right)g_{k+1}(\omega r)C^{(k+1)}_{-q+\lambda} \right. \nonumber
\\
&+& \left. \sqrt{(k+1)(2k-1)}
\left(
\begin{array}{ccc}
1&k-1&k\\
\lambda&q-\lambda&q
\end{array}
\right)g_{k-1}(\omega r)C^{(k-1)}_{-q+\lambda}
 \right]i^{-k-1}(-1)^{k+q-\lambda},
\end{eqnarray}
\begin{eqnarray}\label{2.3.2}
{}_{{\rm l}}A^{(1\lambda)}_{\omega kq}&=&\sqrt{\frac{\omega}{\pi
c(2k+1)}}\left[\sqrt{(k+1)(2k+3)} \left(
\begin{array}{ccc}
1&k+1&k\\
\lambda&q-\lambda&q
\end{array}
\right)g_{k+1}(\omega r)C^{(k+1)}_{-q+\lambda} \right.\nonumber
\\
&+&  \left. \sqrt{k(2k-1)}
\left(
\begin{array}{ccc}
1&k-1&k\\
\lambda&q-\lambda&q
\end{array}
\right)g_{k-1}(\omega r)C^{(k-1)}_{-q+\lambda}
 \right]i^{-k-1}(-1)^{k+q+\lambda+1},
\end{eqnarray}
where $C^{(k)}_{-q}=\sqrt{\frac{4\pi}{2k+1}}Y_{-q}^{(k)}$ and
usual notations for $3j$-symbols are employed. The spherical
components of the transverse magnetic vector potential look as
\begin{eqnarray}\label{2.4}
{}_{{\rm m}}{A}^{(1\lambda)}_{\omega
qk}=(-1)^{\lambda+k+q}i^{-k}\sqrt{\frac{\omega(2k+1)}{\pi
c}}g_k\left(\omega r \right)\left(
\begin{array}{ccc}
1&k&k\\
-\lambda&-q+\lambda&-q
\end{array}
\right)C^{(k)}_{-q+\lambda},
\end{eqnarray}
while the spherical components of the scalar potential are given
by the following expression
\begin{eqnarray}\label{2.5}
\Phi_{\omega kq
}=i^{-k}(-1)^{k+q}2\sqrt{\frac{\omega}{c}}g_k(\omega
r)Y_{-q}^{(k)}.
\end{eqnarray}
The radial functions $g_k(\omega r)$ are related to Bessel
functions $J_\mu(z)$ via
$g_{k}(z)=(2\pi)^{3/2}\frac{1}{\sqrt{z}}J_{k+\frac{1}{2}}(z)$.

Usually two gauges are used: The so-called Coulomb gauge that
corresponds to the vanishing longitudinal part of the vector
potential and the scalar potential (i.e. $\nabla\cdot{\bf
A}=\nabla\cdot {}_{{\rm e}}{\bf A}=0$ and $\Phi =0$). This gauge
is characterized by the choice of the gauge parameter $K=0$.
Another convenient gauge is defined by the following value of the
parameter $K=-\sqrt{\frac{k+1}{k}}$. Within this gauge, as it can
be seen from Eqs.~(\ref{2.3.1}) and (\ref{2.3.2}), the terms
containing spherical functions $C^{(k)}_{-q}$ of the order $k-1$,
vanish in the expression for the transition probability
(\ref{2.1}).

After some manipulations the expression for the probability of
emission of an electric photon with the angular momentum $k$ can
be cast into the form
\begin{eqnarray}\label{2.6}
W_{A\rightarrow A'}^{{\rm
E}k}=\frac{2(k+1)\omega}{k(2k+1)c}\sum\limits_{q=-k}^k\left|\langle
A'|\left[{}_{{\rm
e}}O'^{(k)}_{-q}+K\sqrt{\frac{k}{k+1}}\left({}_{{\rm
l}}O^{(k)}_{-q}+{}_{\Phi}O^{(k)}_{-q}\right)\right]|A\rangle
\right|^2.
\end{eqnarray}
Here
\begin{eqnarray}\label{2.7}
{}_{{\rm
e}}O'^{(k)}_{-q}&=&-i\left[k\sqrt{\frac{2k+3}{k+1}}g_{k+1}(\omega
r
)\left[C^{(k+1)}\times\alpha^{(1)}\right]_{-q}^{(k)}\right.\nonumber
\\
& &+
 \left.\sqrt{k(2k-1)}g_{k-1}(\omega r)\left[C^{(k-1)}\times\alpha^{(1)}\right]_{-q}^{(k)}\right],\\[3pt]
{}_{{\rm l}}O^{(k)}_{-q}&=&i\left[\sqrt{(k+1)(2k+3)}g_{k+1}(\omega
r )\left[C^{(k+1)}\times
\alpha^{(1)}\right]_{-q}^{(k)}\right.\nonumber
\\
& & -
 \left.\sqrt{k(2k-1)}g_{k-1}(\omega r)\left[C^{(k-1)}\times\alpha^{(1)}\right]_{-q}^{(k)}\right], \\[3pt]
{}_{\Phi}O^{(k)}_{-q}&=&\sqrt{2k+1}g_k(\omega r)C^{(k)}_{-q},
\end{eqnarray}
and $[a^{(s_1)}\times b^{(s_2)}]^{(s)}_q$ represents the tensor
product of two irreducible spherical tensors of rank $s_1$ and
$s_2$ coupled to a spherical tensor of rank $s$ with components
$q$.

Using the following integral relation for the Dirac wave functions
\cite{Akhiezer}
\begin{eqnarray}\label{2.8}
i\int\psi^*_{A'}\left({\bf \alpha}{\nabla}\chi\right)\psi_A\,d^3\tau=\frac{\omega}{c}\int\psi^*_{A'}\chi\psi_A
\,d^3\tau,
\end{eqnarray}
where $\chi$ is an arbitrary function, one can establish another
form for the E$k$-transition probability (see \cite{NikRudz}):
\begin{eqnarray}\label{2.9}
W^{{\rm E}k}_{A\rightarrow
A'}=\frac{2(k+1)\omega^3}{k(2k+1)c^3}\sum\limits_{q=-k}^{k}\left|\langle
A'|{}_{{\rm
e}}O_{-q}^{(k)}+K\frac{c}{\omega}\sqrt{\frac{k}{k+1}}\left[{}_{{\rm
l}}O_{-q}^{(k)}+{}_{\Phi}O^{(k)}_{-q}\right]|A\rangle \right|^2,
\end{eqnarray}
where
\begin{eqnarray}\label{2.10}
{}_{{\rm e}}O^{(k)}_{-q}&=&-rg_k(\omega
r)C^{(k)}_{-q}-i\frac{r}{k+1}g_k(\omega r)
\,\left[\sqrt{k(2k-1)}\left[C^{(k-1)}\times\alpha^{(1)}\right]_{-q}^{(k)}\right.
\nonumber \\
&&
+\left.\sqrt{(k+1)(2k+3)}\left[C^{(k-1)}\times\alpha^{(1)}\right]_{-q}^{(k)}\right].
\end{eqnarray}
Thus, we have two different (equivalent) forms for the
E$k$-transition probabilities (Eqs.~(\ref{2.6}) and (\ref{2.9}))
together with an arbitrary choice for the gauge constant $K$ at
our disposal. Analogous expressions for the emission probability
of a photon, characterized by its energy and polarization, are
provided in \cite{KKR}.

The aim of the present investigation concerns the derivation of
the nonrelativistic limit of the E1M1-, E1E1- and E1E2-transition
probabilities in different gauges and forms. Deriving the
nonrelativistic limit of Eqs.~(\ref{2.6}) and (\ref{2.9}), implies
two distinct nonrelativistic forms for the one-photon transition
probability with arbitrary gauge constant $K$ (see \cite{KMR}):
\begin{eqnarray}\label{2.11}
W^{{\rm E}k}_{A\rightarrow
A'}=\frac{2(k+1)(2k+1)\omega^{2k-1}}{k[(2k+1)!!]^2c^{2k+1}}\sum\limits_{q=-k}^k\left|(A'|\left(Q'^{(k)}_{-q}+K\sqrt{\frac{k}{k+1}}\left[Q'^{(k)}_{-q}-\omega
Q^{(k)}_{-q}\right]\right)|A)\right|^2
\end{eqnarray}
and
\begin{eqnarray}\label{2.12}
W^{{\rm E}k}_{A\rightarrow
A'}=\frac{2(k+1)(2k+1)}{k[(2k+1)!!]^2}\sum\limits_{q=-k}^k\left(\frac{\omega}{c}\right)^{2k+1}\left|(A'|\left(Q_{-q}^{(k)}+K\sqrt{\frac{k}{k+1}}\left[\frac{1}{\omega}Q'^{(k)}_{-q}-Q^{(k)}_{-q}\right]\right)|A)\right|^2.
\end{eqnarray}
Here $|A)$ and $(A'|$ are nonrelativistic Schr\"{o}dinger states
(wave functions) together with operators
\begin{eqnarray}\label{2.13}
Q^{(k)}_{-q}&=&-r^kC^{(k)}_{-q},
\end{eqnarray}
\begin{eqnarray}\label{2.14}
Q'^{(k)}_{-q}&=&-r^{-k-1}\left(kC^{(k)}_{-q}\frac{\partial}{\partial
r}+\frac{i}{r}\sqrt{k(k+1)}\left[C^{(k)}\times
L^{(1)}\right]_{-q}^{(k)}\right),
\end{eqnarray}
where $L^{(1)}$ is the orbital angular momentum of the atomic
electron. Choosing $K=0$, we find that the operator in
Eq.~(\ref{2.11}) corresponds to the nonrelativistic transition
operator in the ``velocity'' form, while for
$K=-\sqrt{\frac{k+1}{k}}$ it is related to the transition operator
in the ``length'' form. However, the correspondence of a certain
gauge choice to a particular type of nonrelativistic transition
operators is not unique. In view of Eq.~(\ref{2.12}) we can
conclude, that within the nonrelativistic limit the expression
Eq.~(\ref{2.9}) with $K=0$ converts the transition operator into
the ``length'' form and with $K=-\sqrt{\frac{k+1}{k}}$  into the
``velocity'' form, respectively.

\section{Application of the Coulomb Green Function}

In order to evaluate the transition probabilities for the
processes ${\rm 2p}\rightarrow 2\gamma+{\rm 1s}$ and ${\rm
2s}\rightarrow 2\gamma+{\rm 1s}$ in the hydrogen atom the
nonrelativistic Coulomb Green function is employed. The summations
over the entire spectrum of the Schr\"{o}dinger equation arise
usually when perturbation theory is applied. The Green function
approach allows one to express the intermediate summations in a
closed analytic form. This is very useful for the analysis and for
tests of numerical calculations.

The Green function for the Schr\"{o}dinger equation with the
Hamiltonian $\hat{H}$ is defined by the solution of the equation
\begin{eqnarray}
\left(\hat{H}-E\right)G(E;{\bf r},{\bf r}')=\delta({\bf r}-{\bf
r}')\label{3.20}
\end{eqnarray}
and can be always represented in terms of a spectral decomposition
\begin{eqnarray}\label{spectr}
G(E; {\bf r},{\bf r}')=\sum\limits_{N}\,\frac{\varphi^*_N({\bf
r})\varphi_N({\bf r}')}{E_N-E}.
\end{eqnarray}
In Eq.~(\ref{spectr}) the sum runs over the entire spectrum of the
Hamiltonian (bound and continuous spectrum). The set of quantum
numbers $N$ may be specified as usual by the principal quantum
number $n$, orbital angular momentum number $l$ and projection
$m$. In view of the spherical symmetry it is sufficient to derive
a closed expression for the radial part $g_l(E; r, r')$ of the
Green function defined by the partial wave decomposition
\begin{eqnarray}\label{3.22}
G(E;{\bf r},{\bf
r}')=\sum\limits_{lm}\frac{1}{rr'}\,g_l(E;r,r')\,Y_{lm}^*({\bf
n}_{\bf r})Y_{lm}({\bf n}_{\bf r'}).
\end{eqnarray}
In the particular case of an external Coulomb potential the Green
function of Eq.~(\ref{3.20}) is called Coulomb Green Function
(CGF). With the use of the expansion Eq.~(\ref{3.22}) the radial
integrals occurring in Eqs.~(\ref{2.11}) and (\ref{2.12}) for the
transition probabilities can be calculated explicitly (see Ref.
\cite{GFLabSol} for details).

For the radial part of the Coulomb Green function it is convenient
to employ the Sturmian expansion \cite{RZM}, which is defined in
the entire complex energy plane via
\begin{eqnarray}\label{3.23}
\frac{1}{rr'}\,g_l(E;r,r')=\sum\limits_{n_r=0}^{\infty}\frac{\Phi_{n_rl}(r)\Phi_{n_rl}(r')}{E_{n_rl}-E},
\end{eqnarray}
where $\Phi_{n_rl}(r)$ denote the Sturmian functions. The Sturmian
expansion of the CGF can be represented in an alternative form by
introducing radial functions
\begin{eqnarray}
R_{n_rl}\left(\frac{2r}{\nu}\right)=\frac{1}{r}\sqrt{\frac{Z}{\nu
n_r
}}\frac{1}{(2l+1)!}\sqrt{\frac{\Gamma(n_r+l+1)}{\Gamma(n_r-l)}}M_{n_r,
l+\frac{1}{2}}\left(\frac{2r}{\nu}\right),
\end{eqnarray}
which are related to Whittaker functions $M_{n_r,
l+\frac{1}{2}}\left(\frac{2r}{\nu}\right)$, where
$\nu=Z/\sqrt{-2E}$. For integer values $\nu=n$ these functions
coincide with the normalized, radial hydrogenic wave functions
\begin{eqnarray}\label{3.25}
\Phi_{n_rl}(r)=\sqrt{\frac{\nu
n}{Z}}R_{nl}\left(\frac{2r}{\nu}\right).
\end{eqnarray}
Substitution of Eq.~(\ref{3.25}) into (\ref{3.23}) yields
\begin{eqnarray}
\frac{1}{rr'}\,g_l(\nu;r,r')=\frac{\nu^2}{Z^2}\sum\limits_{n=l+1}^{\infty}\frac{n}{n-\nu}R_{nl}\left(\frac{2r}{\nu}\right)R_{nl}\left(\frac{2r'}{\nu}\right).
\end{eqnarray}
Within this paper we apply the Green function method for the
evaluation of the two-photon decay probability in the hydrogen
atom.

In \cite{E1E1ZMR} the two-photon transition process ${\rm
2s}\rightarrow \gamma({\rm E1})+\gamma({\rm E1})+{\rm 1s}$ has
been considered. The probability for the two-photon decay
$A\rightarrow \gamma({\rm E1})+\gamma({\rm E1})+A'$ with photon
frequencies $\omega_1$ and $\omega_2$ within the nonrelativistic
limit and dipole approximation is:
\begin{eqnarray}\label{3.27}
dW^{\rm E1E1}_{A\rightarrow
A'}(\omega_2)&=&\frac{8}{9\pi}\left(\frac{4\pi}{3}\right)^3\sum\limits_{M_1M_2}\left|(A'|rY_{1M_2}\left({\bf
n}_{\bf r}\right)G(E_{A}-\omega_1;{\bf r},{\bf
r}')r'Y_{1M_1}^*\left({\bf n}_{\bf r'}\right)|A)\right.\nonumber
\\
&& + \left.(A'|rY_{1M_1}\left({\bf n}_{\bf r}\right)G(E_{A}-\omega_2;{\bf r}, {\bf r}\,')r'Y^*_{1M_2}\left({\bf n}_{\bf r'}\right)|A)\right|^2 (\omega_1\omega_2)^2\,d\omega_2\, .
\end{eqnarray}
The energy conservation law implies
$\omega_1=E_A-E_{A'}-\omega_2$. After the evaluation of angular
matrix elements in Eq.~(\ref{3.27}) the remaining radial integrals
have the form
\begin{eqnarray}
\int\limits_{0}^{\infty}\int\limits_{0}^{\infty}\int\limits_{0}^{\infty}dr'drdx\,(r')^{s'+\frac{7}{2}}(r)^{s+\frac{7}{2}}\,\exp\left(-\frac{1}{\nu}(\beta'r'+\beta
r)+(r+r')\cosh(x)\right)\nonumber
\\
\times\left(\coth\left(\frac{x}{2}\right)\right)^{2\nu}\,
I_{2l+1}\left(\frac{2\sqrt{rr'}}{\nu}\sinh(x)\right),
\end{eqnarray}
where $n$ and $n'$ are the principal quantum numbers of the
initial and final states, respectively, together with the
parameters $\beta={\nu}/{n}, \beta'={\nu}/{n'}$ and
$\nu=\sqrt{-2\left(E_{nl}-\omega\right)}$. These integrals can be
evaluated analytically after inserting the series expansion for
the modified Bessel functions $I_{2l+1}$. The integration  over
$x$ should be done at the end. Assuming emission of two E1
photons, one can write the total probability for such a two-photon
decay as
\begin{eqnarray}\label{3.53}
W_{A\rightarrow A'}^{\rm
2E1}=\frac{1}{2}\int\limits_0^{\omega_{0}} dW_{A\rightarrow
A'}^{\rm 2E1}(\omega)
\end{eqnarray}
with $\omega_0 = E_A - E_{A'}$. For the process ${\rm
2s}\rightarrow 2\gamma({\rm E1})+{\rm 1s}$ in \cite{E1E1ZMR} the
result of the evaluations was reported as
\begin{eqnarray}\label{2E1zf}
W_{{\rm 2s}\,{\rm 1s}}^{2E1}=8.226 (\alpha Z)^6\,{\rm s}^{-1}
\end{eqnarray}
with an accuracy of about $1\%$.

However, in further calculations it is convenient to utilize the
other representation of  the radial Coulomb Green function in
terms of an expansion over Laguerre polynomials \cite{RZM}
(employed also in Ref. \cite{E1E1ZMR})
\begin{eqnarray}\label{3.55}
g_l(\nu; r,r')=\frac{4Z}{\nu}\left(\frac{4}{\nu^2}rr'\right)^l
\exp\left(-\frac{r+r'}{\nu}\right)\sum\limits_{n=0}^{\infty}\frac{n!L^{2l+1}_n\left(\frac{2r}{\nu}\right)L^{2l+1}_n\left(\frac{2r'}{\nu}\right)}{(2l+1+n)!(n+l+1-\nu)}.
\end{eqnarray}
The series Eq.~(\ref{3.55}) converges absolutely as $n^{-3/2}$ for
arguments $r, r'>0$ and Im$(\nu)=0$ \cite{RZM}. The angular
momentum quantum number $l=1$ for the intermediate states is fixed
after the angular integration. Inserting the expansion
(\ref{3.55}) for $l=1$ into the expression (\ref{3.27}) yields
\begin{eqnarray}\label{2s.rad.1}
dW_{{\rm 2s}\,{\rm 1s}}^{\rm
2E1}(\omega)=\left(\frac{2}{3}\right)^3\frac{\omega^3\omega'^3}{\pi}\left[I_1(\nu)+I_1(\nu')\right]^2\alpha^6d\omega,
\end{eqnarray}
\begin{eqnarray}\label{2s.rad.2}
I_1(\nu)=\frac{16\sqrt{2}}{\nu^3}\left(\frac{\nu}{2}\right)^{10}\sum\limits_{m=0}^{\infty}\frac{m!}{(m+3)!(m+2-\nu)}\int\limits_0^{\infty}d\xi\xi^4e^{-\xi\left(\frac{\nu+1}{2}\right)}L_m^3(\xi)\nonumber
\\
\times
\int\limits_0^{\infty}dtt^4e^{-t\left(\frac{\nu+2}{4}\right)}\left(1-\frac{\nu}{4}t\right)L_m^3(t).
\end{eqnarray}
These integrals can be evaluated analytically. After inserting
this result into Eq.~(\ref{3.53}) the integration over $\omega$
has to be performed in order to obtain the total transition
probability for the E1E1 decay of the 2s state in the hydrogen
atom. This is achieved numerically with the aid of the
computer-algebra code MATHEMATICA. The final result is
\begin{eqnarray}\label{3.58}
W^{\rm 2E1}_{{\rm 2s}\,{\rm
1s}}=\frac{1}{2}\int\limits_0^{\omega_0}dW^{\rm 2E1}_{{\rm
2s}\,{\rm 1s}}(\omega)=0.00131823\,(\alpha Z)^6 \, {\rm a.u.}
=8.22932 \, {\rm s}^{-1}\, (Z=1)
\end{eqnarray}
with $\omega_0 = E_{\rm 2s} - E_{\rm 1s}$. In Eq.~(\ref{3.58}) we
indicated the Z-dependence of the $W^{\rm E1E1}_{\rm 2s\,1s}$
transition probability. The numerical value (\ref{3.58}) coincides
with most precise result \cite{Jentschura} up to 5 digits. This
will serve us as accuracy estimate of our approach in following
calculations.

\section{E1E2 decay probability for the 2p state}

In this section we consider the E1E2 decay of the 2p state in the
hydrogen atom. Again the set of quantum numbers $nlm_l$ is
employed as far as the total angular momentum $j$ is not important
in this calculation performed within the nonrelativistic approach.
Nevertheless, in order to compare our results with those obtained
from the relativistic evaluation (see Refs. \cite{LabShon},
\cite{LabShonSol}), we shall perform the calculation within two
different gauges according to Eqs.~(\ref{2.11})-(\ref{2.14}). This
will also elucidate the potential influence of relativistic
effects associated with the contribution of the negative-energy
Dirac spectrum.

As a test for the method the gauge constant
$K=-\sqrt{\frac{k+1}{k}}$ is chosen in the expression
Eq.~(\ref{2.11}) for the transition probability, which corresponds
to the nonrelativistic ``length'' form as mentioned above. This
would be equivalent to the choice $K=0$ together with the form
Eq.~(\ref{2.12}). Inspection of Eqs.~(\ref{2.11}) and (\ref{2.13})
reveals that the emission of electric photons (E$k$) is described
by the potentials
\begin{eqnarray}\label{4.1}
V^{{\rm
E}k}=\sqrt{\frac{k+1}{k}}\frac{2\omega^{k+\frac{1}{2}}}{(2k+1)!!}r^kY_{k\,-q}.
\end{eqnarray}
Accordingly, the two electric photon decay rate of the atomic
state $A$ can be written as
\begin{eqnarray}\label{4.2}
dW_{A\rightarrow A'}^{{\rm E}k{\rm E}k'}=\sum\limits_{q
q'm_Am_{A'}}\left|\sum\limits_N\frac{(A'|V^{{\rm E}k}|N)(N|V^{{\rm
E}k'}|A)}{E_N-E_A+\omega}+ \sum\limits_N\frac{(A'|V^{{\rm
E}k'}|N)(N|V^{{\rm E}k}|A)}{E_N-E_A+\omega'}\right|^2\delta
\left(\omega+\omega'-E_A+E_{A'}\right)\, d\omega d\omega'.
\end{eqnarray}
Here the labels $A, A'$ and $N$ abbreviate the set of
nonrelativistic quantum numbers (principal quantum number $n$,
orbital momentum $l$ and projection $m_l$) for indicating the
state of the atomic electron as the initial ($A$), intermediate
($N$) and final ($A'$). The photons will be characterized by the
angular momentum and its projection ($kq$) as well as by the type
of the photon (electric or magnetic). The Eq.~(\ref{4.2}) also
implies the summation over degenerate substates of the final
atomic state $A'$ and the average over the degenerate substates of
the initial atomic state $A$ as well as  summations over the
angular momentum projections of both emitted photons. The
frequencies of the two photons $\omega$ and $\omega'$ are related
by the energy conservation law $\omega'=\omega_0-\omega$, where
$\omega_0=E_A-E_{A'}$.

Employing the eigenmode decomposition of the Coulomb Green
function Eq.~(\ref{3.22}) the probability of the two-photon decay
process takes the form
\begin{eqnarray}\label{4.3}
dW^{{\rm E}k{\rm E}k'}_{A\rightarrow
A'}=\frac{2\pi}{2l_A+1}\sum\limits_{qq' m_A
m_{A'}}\left|\sum\limits_{lm_l}\int\int d{\bf r_1}d{\bf
r_2}\,R_{n_{A'}l_{A'}}(r_1)Y^*_{l_{A'}m_{A'}}({\bf n}_{\bf
r_1})\right.\nonumber
\\
\nonumber
 \left.\times V^{{\rm E}k}({\bf
 r_1})g_l(\nu;r_1,r_2)Y_{lm_l}({\bf n}_{\bf r_1})Y^*_{lm_l}({\bf
 n}_{\bf r_2})V^{{\rm E}k'}({\bf r_2})R_{n_Al_A}(r_2)Y_{l_Am_A}({\bf n}_{\bf r_2})+\right.
\\
\nonumber
 \left.+\sum\limits_{lm_l}\int\int d{\bf r_1}d{\bf r_2}\,R_{n_{A'}l_{A'}}(r_1)Y^*_{l_{A'}m_{A'}}({\bf n}_{\bf r_1})
V^{{\rm E}k'}({\bf r_1})g_l(\nu';r_1, r_2)Y_{lm_l}({\bf n}_{\bf
r_1}) \right.
\\
 \left.
\times Y^*_{lm_l}({\bf n}_{\bf r_2})V^{{\rm E}k}({\bf
r_2})R_{n_Al_A}(r_2)Y_{l_Am_A}({\bf n}_{\bf r_2})
 \right|^2d\omega,\nonumber \\
\end{eqnarray}
where $V^{{\rm E}k}({\bf r})$ is the potential (\ref{4.1}) written
in the gauge $K=-\sqrt{\frac{k+1}{k}}$ and compatible with the
form Eq.~(\ref{2.11}), together with parameters
$\nu=Z/\sqrt{-2(E_A-\omega)}$, $\nu'=Z/\sqrt{-2(E_A-\omega')}$ and
frequency $\omega'=E_A-E_{A'}-\omega$.

Specifying Eq.~(\ref{4.3}) for the transition between levels
$A={\rm 2p}$, $A'={\rm 1s}$ and taking into account that in this
case the angular momentum of the photon can take values $k=1, 2$,
we receive four different terms contributing in Eq.~(\ref{4.3}).
Unfortunately, in the previous paper \cite{LabSol} an error in the
summation over projections of the Clebsch-Gordan coefficients did
occur (see expression Eq.~(41) in \cite{LabSol}). Here we correct
this mistake and give the proper expression for the probability.
After angular integration and summation over projections we find
\begin{eqnarray}\label{4.4}
dW_{{\rm 2p}\,{\rm 1s}}^{\rm
E1E2}(\omega)=\frac{2^2\omega^3\omega'^3}{3^35^2\pi}\left[\omega'^2\left|I_1(\omega')+I_2(\omega)\right|^2+\omega^2
\left|I_1(\omega)+I_2(\omega')\right|^2\right]d\omega,
\end{eqnarray}
where
\begin{eqnarray}\label{4.5}
I_1(\omega)=\frac{1}{\sqrt{6}}\int\limits_0^{\infty}\int\limits_0^{\infty}dr_1dr_2\,r_1^3r_2^5\,e^{-r_1-\frac{r_2}{2}}\,g_1(E_A-\omega;
r_1, r_2)
\end{eqnarray}
and
\begin{eqnarray}\label{4.6}
I_2(\omega)=\frac{1}{\sqrt{6}}\int\limits_0^{\infty}\int\limits_0^{\infty}dr_1dr_2\,r_1^4r_2^4\,e^{-r_1-\frac{r_2}{2}}\,g_2(E_A-\omega;
r_1, r_2),
\end{eqnarray}
respectively. Inserting again the representation Eq.~(\ref{3.55})
for the CGF with $l=1$ leads to radial integrals that can be
evaluated analytically.

Substituting the integrals (\ref{4.5}) and (\ref{4.6}) into
(\ref{4.4}) and integrating over frequencies $\omega$ yields
\begin{eqnarray}\label{4.9}
W^{\rm E1E2}_{{\rm 2p}\,{\rm
1s}}=\frac{1}{2}\int\limits_0^{\omega_0} dW_{{\rm 2p}\,{\rm
1s}}^{\rm E1E2} =1.98896\cdot 10^{-5}\,(\alpha Z)^8\, {\rm a.u.} =
6.61197\cdot 10^{-6}\, {\rm s}^{-1}\,(Z=1)
\end{eqnarray}
with $\omega_0=E_{\rm 2p}-E_{\rm 1s}$. In Eq.~(\ref{4.9}) we
indicated the Z-dependence of the $W^{\rm E1E1}_{\rm 2p\,1s}$
transition probability. Compared with the relativistic result in
the ``length'' gauge (see \cite{LabShonSol}) the relative
discrepancy is about $0.1\%$.

The calculation of the E1E2 two-photon decay with the
nonrelativistic ``velocity'' form is more involved. Now the gauge
constant should be chosen either $K=-\sqrt{\frac{k+1}{k}}$ for the
form Eq.~(\ref{2.12}) or $K=0$ for the form Eq.~(\ref{2.11}).

We choose $K=0$ together with the form Eq.~(\ref{2.11}). The potential in this case reads
\begin{eqnarray}\label{4.10}
V^{{\rm E}k}({\bf
r}\,)=\frac{4\omega^{k-\frac{1}{2}}}{(2k+1)!!}\sqrt{\frac{k+1}{k(2k+1)}}r^{k-1}\left[kY^{(k)}_{-q}({\bf
n}_{\bf r})\frac{\partial}{\partial
r}+\frac{i}{r}\sqrt{k(k+1)}\left[Y^{(k)}\times
L^{(1)}\right]^k_{-q}\right].
\end{eqnarray}
The formula for $dW_{{\rm 2p}\,{\rm 1s}}^{\rm E1E2}$ follows again
from Eq.~(\ref{4.3}). Performing angular integrations and
summations over projections as discussed in previous cases now
yields
\begin{eqnarray}\label{4.11}
dW_{{\rm 2p}\,{\rm 1s}}^{\rm E1E2}(\omega)=
\frac{2^4}{3^35^2\pi}\omega'\omega\left[\omega^2\left|I_1(\omega)+I_2(\omega')\right|^2+\omega'^2\left|I_1(\omega')+I_2(\omega)\right|^2\right]d\omega
\end{eqnarray}
with radial integrals of the type
\begin{eqnarray}\label{4.12}
I_1(\omega)=\frac{1}{\sqrt{6}}\int\limits_0^{\infty}\int\limits_0^{\infty}dr_1dr_2\,
r_1^2r_2^3\,e^{-r_1-\frac{r_2}{2}}\left[1-\frac{9i}{2}-\frac{r_2}{2}\right]\left[\frac{\partial}{\partial
r_1}-\frac{2i}{r_1}\right]\,g_1(E_A-\omega; r_1, r_2)
\end{eqnarray}
and
\begin{eqnarray}\label{4.13}
I_2(\omega)=\frac{1}{\sqrt{6}}\int\limits_0^{\infty}\int\limits_0^{\infty}dr_1dr_2\,r_1^3r_2^2\,
e^{-r_1-\frac{r_2}{2}}\left[1-5i-\frac{r_2}{2}\right]\left[\frac{\partial}{\partial
r_1}-\frac{3i}{r_1}\right]\,g_2(E_A-\omega; r_1, r_2)
\end{eqnarray}
together with parameters $\nu=Z/\sqrt{-2(E_{\rm 2p}-\omega)}$ and
$\nu'=Z/\sqrt{-2(E_{\rm 2p}-\omega')}$, respectively. The
integrations over $r_1$ and $r_2$ lead to a rather lengthy
analytical expression containing various combinations of notations
similar to those in Eqs.~(\ref{4.5}) and (\ref{4.6}). The
numerical evaluation yields finally
\begin{eqnarray}\label{4.14}
W_{{\rm 2p}\,{\rm 1s}}^{\rm
E1E2}=\frac{1}{2}\int\limits_0^{\omega_0}dW_{{\rm 2p}\,{\rm
1s}}^{\rm E1E2} (\omega) = 3.6896\cdot 10^{-6}\,(\alpha Z)^8 \,
{\rm a.u.}\simeq 1.227\cdot 10^{-6}\, {\rm s}^{-1}\, (Z=1),
\end{eqnarray}
where $\omega_0=3/8$ a.u. This result differs from that obtained
from relativistic calculations \cite{LabShon}, \cite{LabShonSol}
by about $0.5\%$. This discrepancy is larger than the relativistic
corrections (of the relative order $\alpha^2$) and can be
attributed to the pure numerical uncertainties in both
calculations. Note, that unlike the case of the ``length'' form,
the negative-energy contribution is no longer negligible when the
``velocity'' form is employed. Therefore, the result
Eq.~(\ref{4.14}) does not coincide with Eq.~(\ref{4.9}) and
represents only the positive-energy contribution to $W^{\rm
E1E2}_{\rm 2p\,1s}$ in the ``velocity'' form. Correspondingly, we
compare this result to the positive-energy contribution calculated
in \cite{LabShon}, \cite{LabShonSol}. The negative-energy
contribution to $W^{\rm E1E2}_{\rm 2p\,1s}$ in the ``velocity''
form for low $Z$ values was evaluated analytically in
\cite{LabShonSol}.

\section{E1M1 two-photon decay}

For the mixed E1M1 two-photon transition probability the
expression (\ref{4.2}) should be replaced by
\begin{eqnarray}\label{5.1}
dW_{A\rightarrow A'}^{\rm
E1M1}=\sum\limits_{M_eM_mm_Am_{A'}}\left|\sum\limits_{N}\frac{(A'|V^{\rm
E1}(\omega)|N)(N|V^{{\rm
M1}}(\omega')|A)}{E_N-E_A+\omega}\right.\nonumber
\\
\nonumber \left. +\frac{(A'|V^{\rm M1}(\omega')|N)(N|V^{\rm
E1}(\omega)|A)}{E_N-E_A+\omega'}\right.
\\
\nonumber
 \left.
+\frac{(A'|V^{\rm E1}(\omega')|N)(N|V^{\rm
M1}(\omega)|A)}{E_N-E_A+\omega'}\right.
\\
 \left.  +\frac{(A'|V^{\rm M1}(\omega)|N)(N|V^{\rm E1}(\omega')|A)}{E_N-E_A+\omega}\right|^2\,d\omega.
\end{eqnarray}
Here $V^{\rm E1}(\omega)=\frac{4}{3}\omega^{3/2}rY_{M_{\rm
e}}^{(1)}$, $V^{\rm
M1}(\omega)=\sqrt{\frac{4}{3}}\mu_0\omega^{3/2}\left(\hat{j}_{1M_{\rm
m}}+\hat{s}_{1M_{\rm m}}\right)$, $Y_{M_e}^{(1)}$ is the spherical
tensor of the rank 1 with the spherical component $M_e$ (i.e. the
spherical function $Y_{1M_e}$), $\mu_0=\frac{\alpha}{2}$ is the
Bohr's magneton, $\hat{j}_{1M_m}$ and $\hat{s}_{1M_m}$ are the
spherical components of the total angular momentum and the spin
operator (spherical tensors of rank 1) of the electron. This
choice corresponds to the nonrelativistic ``length'' form for
describing the emission of the electric photons. Since the
potential for the magnetic photon includes total angular momentum
and spin operator, coupled wave functions with the set of quantum
numbers $N=\{nlsjm\}$ should be used, i.e.
\begin{eqnarray}\label{5.2}
\phi_{nlsjm}=\sum\limits_{m_lm_s}C^{jm}_{lm_l\,sm_s}R_{nl}(r)Y_{m_l}^{(l)}({\bf
n}_{\bf r})\chi_{sm_s},
\end{eqnarray}
where $R_{nl}(r)$ is the solution of the radial Schr\"{o}dinger
equation and $\chi_{sm_s}$ $(s=1/2)$ is the spin function. The
magnetic potentials in Eq.~(\ref{5.1}) do not depend on radial
variables. Thus, only the intermediate state with $nl=n_Al_A$ or
$nl=n_{A'}l_{A'}$ will contribute to the probability in
Eq.~(\ref{5.1}). After performing angular integrations and
summations over all projections one arrives  at the expressions
\begin{eqnarray}\label{5.3}
dW_{{\rm 2p}\,{\rm 1s}}^{\rm
E1M1}(\omega)=\frac{2^8\mu_0^2}{\pi}\left(\frac{2}{3}\right)^{12}\omega\omega'^3d\omega
\end{eqnarray}
and
\begin{eqnarray}\label{5.4}
W_{{\rm 2p}\,{\rm 1s}}^{\rm E1M1}=\frac{1}{2}\int\limits_0^{3/8}
dW_{{\rm 2p}\,{\rm 1s}}^{\rm E1M1}(\omega)=
\frac{2^5}{\pi}\left(\frac{2}{3}\right)^{12}\alpha^8\int\limits_0^{3/8}\omega\left(\frac{3}{8}-\omega\right)^{3}d\omega.
\end{eqnarray}
As the final result we obtain
\begin{eqnarray}\label{5.5}
W_{{\rm 2p}\,{\rm 1s}}^{\rm
E1M1}=\frac{2^5}{\pi}\left(\frac{2}{3}\right)^{12}\frac{243}{655360}(\alpha
Z)^8\, {\rm a.u.} = 9.6769\cdot 10^{-6}\, {\rm s}^{-1}\,(Z=1).
\end{eqnarray}
Again the Z-dependence of the $W^{\rm E1M1}_{\rm 2p\,1s}$
transition probability is indicated. Comparison with the result of
a fully relativistic calculation now reveals a discrepancy of
about $0.1\%$.

\section{E1E2 and E1M1 transition probabilities for the 2p hydrogenic state expressed through the photon momentum and polarization.}

In this section we are going to derive expressions which allow the
analysis of the dependence on the directions and polarization
degrees of freedom of the emitted photons. For this purpose we
turn back to the generic $S$-matrix formulation employing the set
of quantum numbers defined by polarization vector ${\bf e}$ and
wave vector ${\bf k}$. Relativistic units are used throughout this
section.

The $S$-matrix element of the two-photon decay process $A\to
A'+2\gamma$ for the noninteracting electrons is represented by
\begin{eqnarray}\label{u7}
S^{(2\gamma)}_{A'A}=(-i)^2e^2\int
dx_1dx_2\,\left(\bar\psi_{A'}(x_1)\gamma^{\mu}A^*_{\mu}(x_1)S(x_1,x_2)\gamma^{\nu}A^*_{\nu}(x_2)\psi_A(x_2)\right),
\end{eqnarray}
where $x_1=({\bf r}_1,t_1)$ and $x_2=({\bf r}_2,t_2)$ are
4-vectors. Also
\begin{eqnarray}\label{u8}
S(x_1,x_2)=\frac{1}{2\pi
i}\int\limits_{-\infty}^{\infty}d\omega_1\,e^{i\omega_1(t_1-t_2)}\sum_n
\frac{\psi_n({x_1}) \bar\psi_n(x_2)}{E_n(1-i0)+\omega_1}
\end{eqnarray}
is the electron propagator where the sum runs over the Dirac
spectrum for the electron in the field of the nucleus, $\psi_n(x)$
is the electron Dirac wave function, $E_n$ is the electron energy,
\begin{eqnarray}\label{u10}
A^{{\bf k},\lambda}_{\mu}(x)=\sqrt{\frac{2\pi}{\omega}}e^{(\lambda)}_{\mu}e^{i({\bf k}{\bf r}-\omega t)}
\end{eqnarray}
is the wave function of the photon characterized by the momentum
${\bf k}$ and polarization vector $e_{\mu}^{\lambda}$
($\mu,\lambda=1,2,3,4$), $x\equiv({\bf r},t)$. For the real
transverse photons
\begin{equation}\label{u28}
{\bf A}(x)=\sqrt{\frac{2\pi}{\omega}}{\bf e}e^{i({\bf k}{\bf
r}-\omega t)}\equiv\sqrt{\frac{2\pi}{\omega}}{\bf A}_{{\bf e},{\bf
k}}({\bf r})e^{-i\omega t}.
\end{equation}
Inserting Eqs.~(\ref{u8})-(\ref{u28}) in Eq.~(\ref{u7}),
integrating over time and frequency variables and introducing the
amplitude $U_{A'A}$ as
\begin{equation}
S_{AA'}^{(2\gamma)}=-2\pi
i\delta(E_{A'}+\omega+\omega'-E_A)U_{A'A}^{(2\gamma)}
\end{equation}
we obtain
\begin{equation}\label{u11}
U^{(2\gamma)}_{A'A}=\frac{2\pi
e^2}{\sqrt{\omega\omega'}}\sum_n\frac{({\bf \alpha}{\bf A}^*_{{\bf
e},{\bf k}})_{A'n}({\bf \alpha}{\bf A}^*_{{\bf e}',{\bf
k}'})_{nA}}{E_n-E_A+\omega'},
\end{equation}
where $e$ is the electron charge.

Taking into account photon permutation symmetry we define the
transition probability as
\begin{eqnarray}\label{u12}
dW_{A'A}=2\pi\delta(E_A-E_{A'}-\omega-\omega')\left|U^{(2\gamma)a}_{A'A}+U^{(2\gamma)b}_{A'A}\right|^2\frac{d{\bf
k}}{(2\pi)^3}\frac{d{\bf k}'}{(2\pi)^3}.
\end{eqnarray}
Taking $d{\bf k}\equiv\omega^2d{\bf n}_{\bf k}d\omega$ and
integrating over $\omega$ yields
\begin{eqnarray}\label{u13}
dW_{A'A}^{(2\gamma)}(\omega',{\bf n}_{\bf k},{\bf n}_{\bf k'},{\bf
e},{\bf e}')=e^4
\frac{\omega'(E_A-E_A'-\omega')}{(2\pi)^3}\sum\limits_{m_{A}m_{A'}}\frac{1}{2j_{A}+1}\nonumber
\\
\times\left|\sum_{n}\frac{({\bf \alpha}{\bf A}^*_{{\bf e}, {\bf
k}})_{A'n} ({\bf \alpha}{\bf A}^*_{{\bf e}', {\bf
k}'})_{nA}}{E_n-E_A+\omega'}+\sum_{n}\frac{({\bf \alpha}{\bf
A}^*_{{\bf e}', {\bf k}'})_{A'n} ({\bf \alpha}{\bf A}^*_{{\bf e},
{\bf k}})_{nA}}{E_n-E_A+\omega}\right|^2d{\bf n}_{\bf k}{\bf
n}_{\bf k'}d\omega'.
\end{eqnarray}
Here the summation over $n$ abbreviates the summation over the
whole set of quantum numbers $ nj_nl_nm_n$ of intermediate states.
The sums over the projections of the total angular momentum  of
the final state $A'$ and the averaging over the projections of the
total angular momentum of the initial state $A$ in Eq.~(\ref{u13})
are also included.

In the Pauli approximation (see Eq.~(\ref{1.6})) the (\ref{u13})
expression can be rewritten as
\begin{eqnarray}\label{1.this}
dW_{A'A}^{{\rm
P}\,(2\gamma)}=e^4\frac{\omega'(E_A-E_{A'}-\omega')}{(2\pi)^3}\sum\limits_{m_{A}m_{A'}}\frac{1}{2j_{A}+1}\nonumber
\\
\times\left|\sum\limits_n\frac{U^{\rm P}_{A'n}({\bf e},{\bf
k})U^{\rm P}_{nA}({\bf e'},{\bf
k'})}{E-n-E_A+\omega'}+\sum\limits_n\frac{U^{\rm P}_{A'n}({\bf
e'},{\bf k'})U^{\rm P}_{nA}({\bf e},{\bf
k})}{E-n-E_A+\omega}\right|^2d{\bf n}_{\bf k}{\bf n}_{\bf
k'}d\omega'
\end{eqnarray}
where $U^{\rm P}_{A'n}({\bf e},{\bf k})$ is defined by
Eq.~(\ref{1.6}).

Within the Pauli approximation we have to take into account the
two terms of the exponent expansion in the expression for the
emission operator
\begin{eqnarray}\label{E1E2decomp}
A^{\rm P}({\bf e}, {\bf k})=(({\bf e}^*\hat{\bf p})+i({\bf
e}^*[{\bf k}\times\hat{\bf s}]))e^{-i{\bf k}{\bf r}}\approx
\frac{im}{\hbar}[\hat{H},{\bf e}^*{\bf
r}]+\frac{m}{2\hbar}[\hat{H},({\bf e}^*{\bf r})({\bf k}{\bf
r})]+\frac{i}{2}\left({\bf e}^*[{\bf k}\times([{\bf
r}\times\hat{\bf p}]+2\hat{\bf s})]\right).
\end{eqnarray}
The first term in this expression represents the electric dipole
moment of the emitted photon, the second one is the electric
quadrupole moment, and the last term in Eq.~(\ref{E1E2decomp})
represents the magnetic dipole moment of the emitted photon.

For the beginning we evaluate the two-photon E1 and E2 decay rate.
The probability of this process can be written as
\begin{eqnarray}\label{E1E2.3}
dW^{\rm (E1E2)}_{A'A}(\omega')=e^4
\frac{\omega'(E_A-E_{A'}-\omega')}{(2\pi)^3}\sum\limits_{m_{A}m_{A'}}\frac{1}{2j_{A}+1}\nonumber
\\
\times\left|\sum_{n}\frac{U^{\rm P(E1)}_{A'n}({\bf e},{\bf k})
U^{\rm P(E2)}_{nA}({\bf e'},{\bf k'})+U^{\rm P(E2)}_{A'n}({\bf
e},{\bf k}) U^{\rm P({\rm E1})}_{nA}({\bf e'},{\bf
k'})}{E_n-E_A+\omega'}\right.\\
\nonumber\left. +\sum_{n}\frac{U^{\rm P(E1)}_{A'n}({\bf e'},{\bf
k'}) U^{\rm P(E2)}_{nA}({\bf e},{\bf k})+U^{\rm P(E2)}_{A'n}({\bf
e'},{\bf k'}) U^{\rm P(E1)}_{nA}({\bf e},{\bf
k})}{E_n+E_{A'}+\omega'}\right|^2d{\bf n}_{\bf k}{\bf n}_{\bf
k'}d\omega',
\end{eqnarray}
the notations $U^{\rm P(E1)}_{A'n}$ and $U^{\rm P(E2)}_{A'n}$ for
the dipole and quadrupole photons are used in accordance with
decomposition Eq.~(\ref{E1E2decomp}). The Hamiltonian $\hat{H}$ in
Eq.~(\ref{E1E2decomp}) acts on the eigenfunctions and, therefore,
we can rewrite the expression Eq.~(\ref{E1E2.3}) in the form
\begin{eqnarray}\label{E1E2.4}
dW^{\rm (E1E2)}_{A'A}(\omega')=e^4
\frac{\omega'^3\omega^3}{4(2\pi)^3}\sum\limits_{m_{A}m_{A'}}\frac{1}{2j_{A}+1}\nonumber
\\
\times\left|\omega'\sum_{n}\frac{({\bf e}^*{\bf r})_{A'n}(({\bf
e}\,'^*{\bf r})({\bf n}_{\bf k'}{\bf r}))_{nA}}{E_n-E_A+\omega'} +
\omega\sum_{n}\frac{(({\bf e}^*{\bf r})({\bf n}_{\bf k}{\bf
r}))_{A'n}(({\bf e}\,'^*{\bf r}))_{nA}}{E_n-E_A+\omega'}\right. \\
\left.\nonumber +\omega\sum_{n}\frac{({\bf e}\,'^*{\bf
r})_{A'n}(({\bf e}^*{\bf r})({\bf n}_{\bf k}{\bf
r}))_{nA}}{E_n-E_A+\omega} + \omega'\sum_{n}\frac{(({\bf e}'^*{\bf
r})({\bf n}_{\bf k'}{\bf r}))_{A'n}(({\bf e}^*{\bf
r}))_{nA}}{E_n-E_A+\omega} \right|^2d{\bf n}_{\bf k}{\bf n}_{\bf
k'}d\omega',
\end{eqnarray}
where the relation $\omega=E_A-E_{A'}-\omega'$ holds.

For the summation over polarizations we use again
Eq.~(\ref{1.Fik}) and for integrating over the directions of the
emitted photons we employ the formulas
\begin{eqnarray}\label{E1E2.6}
\int d{\bf n}\,n_i=\int d{\bf n}\,n_i n_k
n_l=0,\,\, \int d{\bf n}\,n_i n_k=\frac{4\pi}{3}\delta_{ik},
\\
\nonumber \int d{\bf n}\,n_i n_j n_k
n_l=\frac{4\pi}{15}(\delta_{ij}\delta_{kl}+\delta_{ik}\delta_{jl}+\delta_{il}\delta_{jk}).
\end{eqnarray}
The formula
\begin{eqnarray}\label{E1E2.pol}
\sum\limits_{{\bf e}\, {\bf e'}}|{\bf e}{\bf e'}|^2=1+({\bf
n}_{\bf k}{\bf n}_{\bf k'})^2= 2-[{\bf n}_{\bf k}\times{\bf
n}_{\bf k'}]^2
\end{eqnarray}
is also useful for calculations. We should stress that in our
paper \cite{LabShonSol} an unfortunate misprint does exist:
Eq.~(51) should contain a vector product.

For the evaluation of the quadrupole matrix elements in
Eq.~(\ref{E1E2.4}) it is convenient to make use of the identity
\begin{eqnarray}\label{lastAngInt}
Y_{l_1m_1}({\bf n})Y_{l_2m_2}({\bf
n})=\sum\limits_{LM}\sqrt{\frac{(2l_1+1)(2l_2+1)}{4\pi(2L+1)}}C_{l_1
0\,l_2 0}^{L0}C_{l_1 m_1\,l_2 m_2}^{LM}Y_{LM}({\bf n}).
\end{eqnarray}
Therefore,
\begin{eqnarray}\label{E1E2.7}
({\bf e}^*{\bf r})({\bf n}_{\bf k}{\bf
r})=\sqrt{\frac{2}{3}}\sqrt{\frac{4\pi}{5}}\sum\limits_{p_1p_2M}(-1)^{p_1+p_2}C^{1\,
1}_{2\, \bar{M}}(p_1p_2)C_{1 p_1\,1 p_2}^{2\bar{M}}e^*_{p_1}({\bf
n}_{\bf k})_{p_2}r^2Y_{2M}({\bf n}_{\bf r}).
\end{eqnarray}
In this case the integration over angles can be provided by the
(see \cite{Varsh}) standard relations
\begin{eqnarray}
\label{AngularInt} \int\limits_{4\pi}d{\bf n}\, Y^*_{l_1m_1}({\bf
n})Y_{l_2m_2}({\bf n})=\delta_{l_1l_2}\delta_{m_1m_2},
\end{eqnarray}
\begin{eqnarray}\label{AngularInt.l}
\int\limits_{4\pi}d{\bf n}\, Y^*_{l_1m_1}({\bf n})Y_{l_2m_2}({\bf
n})Y_{l_3m_3}({\bf n})=
\sqrt{\frac{(2l_2+1)(2l_3+1)}{4\pi(2l_1+1)}}C_{l_2 0\,l_3
0}^{l_10}C_{l_2 m_2\,l_3 m_3}^{l_1m_1}.
\end{eqnarray}
Finally, integrating Eq.~(\ref{E1E2.4}) over angles and summing
over all projections of the angular momenta, we receive
\begin{eqnarray}
\label{E1E2.8} dW^{\rm (E1E2)}_{\rm
2p\,1s}(\omega')=e^4\frac{\omega'^3\omega}{(2\pi)^35^23^2}\int\limits_{4\pi}
d{\bf n}_{\bf k}\int\limits_{4\pi} d{\bf n}_{\bf
k'}\sum\limits_{{\bf e}{\bf e'}}|{\bf e}{\bf
e'}|^2\left[\omega'^2[I_1(\omega')+I_2(\omega)]^2+\omega^2[I_1(\omega)+I_2(\omega')]^2\right]d\omega',
\end{eqnarray}
where $I_1(\omega)$, $I_2(\omega)$ are defined by
Eqs.~(\ref{4.5}), (\ref{4.6}). After the summation over
polarizations and integrating over photon directions the previous
result Eqs.~(\ref{4.4}) and (\ref{4.9}) are recovered.

The next step is the evaluation of E1M1 two-photon transition
${\rm 2p}\rightarrow {\rm 1s}+\gamma({\rm M1})+\gamma({\rm E1})$
with the set of quantum numbers of ${\bf e}$ and ${\bf k}$.
Formula Eq.~(\ref{E1E2.3}) can be cast into the form:
\begin{eqnarray}\label{E1E2.10}
dW^{\rm (E1M1)}_{A'A}(\omega')=e^4
\frac{\omega'(E_A-E_{A'}-\omega')}{(2\pi)^3}\sum\limits_{m_{A}m_{A'}}\frac{1}{2j_{A}+1}\nonumber
\\
\times\left|\sum_{n}\frac{U^{\rm P(E1)}_{A'n}({\bf e},{\bf k})
U^{\rm P(M1)}_{nA}({\bf e'},{\bf k'})+U^{\rm P(M1)}_{A'n}({\bf
e},{\bf k}) U^{\rm P(E1)}_{nA}({\bf e'},{\bf
k'})}{E_n-E_A+\omega'}\right.\\
\nonumber\left. +\sum_{n}\frac{U^{\rm P(E1)}_{A'n}({\bf e'},{\bf
k'}) U^{\rm P(M1)}_{nA}({\bf e},{\bf k})+U^{\rm P(M1)}_{A'n}({\bf
e'},{\bf k'}) U^{\rm P(E1)}_{nA}({\bf e},{\bf
k})}{E_n+E_{A'}+\omega'}\right|^2d{\bf n}_{\bf k}{\bf n}_{\bf
k'}d\omega'
\end{eqnarray}
and, according to Eq.~(\ref{E1E2decomp}),
\begin{eqnarray}\label{E1E2.11}
dW^{\rm (E1M1)}_{A'A}(\omega')=e^4
\frac{\omega'\omega}{4(2\pi)^3}\frac{1}{2j_{A}+1}\sum\limits_{m_{A}m_{A'}}\nonumber
\\
\times\left|\omega\omega'\sum_{n}\frac{({\bf e}^*{\bf r})_{A'n}
\left({\bf e}\,'^*[{\bf n}_{\bf k'}\times (\hat{\bf l}+2\hat{\bf
s})]\right)_{nA}}{E_n-E_A+\omega'}+\omega\omega'\frac{\left({\bf
e}^*[{\bf n}_{\bf k}\times (\hat{\bf l}+2\hat{\bf
s})\right)_{A'n} ({\bf e}'^*{\bf r})_{nA}}{E_n-E_A+\omega'}\right.\\
\nonumber\left. +\omega\omega'\sum_{n}\frac{({\bf e}\,'^*{\bf
r})_{A'n} \left({\bf e}^*[{\bf n}_{\bf k}\times (\hat{\bf
l}+2\hat{\bf s})]\right)_{nA}}{E_n-E_A+\omega}+\omega\omega'
\frac{\left({\bf e}'^*[{\bf n}_{\bf k'}\times (\hat{\bf
l}+2\hat{\hat s})]\right)_{A'n} ({\bf e}^*{\bf r})_{nA}({\bf
e}{\bf k})}{E_n-E_A+\omega}\right|^2d{\bf n}_{\bf k}{\bf n}_{\bf
k'}d\omega',
\end{eqnarray}
where $\hat{\bf l}=[{\bf r}\times\hat{\bf p}]$.

All the matrix elements can be easily evaluated with the aid of
Eqs.~(\ref{AngularInt})-(\ref{AngularInt.l}). Finally, we arrive
again at Eqs.~(\ref{5.3})-(\ref{5.5}):
\begin{eqnarray}\label{E1E2.12}
dW^{\rm (E1M1)}_{\rm 2p\,1s}=
e^4\frac{2^5}{\pi}\left(\frac{2}{3}\right)^{12}(3/8-\omega')\omega'\left((3/8-\omega')^2+\omega'^2\right)d\omega',
\end{eqnarray}
\begin{eqnarray}
W_{\rm 2p\,1s }^{{\rm E1M1}}= \frac{1}{2}\int\limits_0^{3/8}
dW_{2p\,1s}^{{\rm E1M1}}= \frac{1}{\pi}\frac{1}{10935}(\alpha
Z)^8\, {\rm a.u.} = 9.6769\cdot 10^{-6}\, {\rm s}^{-1}\,(Z=1) .
\end{eqnarray}
Thus, in this section we have evaluated again the two-photon
transition probabilites, E1E2 and E1M1, for the process of 2p
state decay in hydrogen atom. In contrast to \cite{LabSol},
calculations were performed based on the representation dealing
with the other set of quantum numbers. The analytic results
obtained are in a good agreement with corresponding relativistic
values (discrepancy is not more then $0.1\%$). Recently the two-photon emission probabilities was evaluated by fully numerical methods in \cite{Amaro}. The corresponding results are in perfect agreement. The method applied
in this section simplifies the determination  of the dependence on
directions of the emitted photons in the two-photon transition
probability when the external field is present.

\section{Two-photon decay of the 2s and 2p hydrogenic states in an external electric field}

In this part of our work we evaluate the two-photon 2s-1s
transition probability for the hydrogenic atoms in the presence of
an external electric field. Similar as in section II of this paper
we take into account only the mixing of the 2s and 2p states. Such
level mixing leads to additional E1E2 and E1M1 two-photon decays
besides the dominant "pure" E1E1 two-photon transition. As it was
shown in previous sections the E1E2 and E1M1 two-photon transition
probabilities are about $(\alpha Z)^2$ times smaller then E1E1
transition probability. However, a consideration must be given to
the fast increasing accuracy of the spectroscopical experiments.
Therefore, E1E2 and E1M1 two-photon transition rates should be
taken into account like a correction to E1E1 decay. Moreover, we
will show that in the presence of the external electric field
terms linear in the field will add to E1E1 transition probability;
These are the interference terms. Furthermore, most if not all
spectroscopical experiments (see most accurate experiments
\cite{4,5}) involve an external electric field so the analysis of
the two-photon decay transitions in an external electric field is
indeed required.

In order to evaluate the two-photon $\overline{\rm 2s}\rightarrow
{\rm 1s}+2\gamma$ transition probability in an external electric
field we should turn to Eq.~(\ref{1.this}) and use the
decomposition Eq.~(\ref{E1E2decomp}). The mixing of the 2s and
$2p$ states is described by Eq.~(\ref{1.mixing}). According to
Eq.~(\ref{DmElem}), the two-photon transition probability in an
external weak electric field can be introduced in the form:
\begin{eqnarray}\label{1.extern}
dW_{A'A}^{(2\gamma)}(\omega',{\bf n}_{\bf k},{\bf n}_{\bf k'},
{\bf e},{\bf e'})=e^4
\frac{\omega'\omega}{(2\pi)^3}\sum\limits_{\mu'\mu''}\frac{1}{2j''+1}\Big|
\sum_{n}\frac{\langle {\rm 1s}\mu'|A^{\rm P}({\bf e},{\bf
k})|n\rangle\langle n|A^{\rm P}({\bf e'},{\bf k'})|{\rm
2s}\mu''\rangle}{E_n-E_A+\omega'} \nonumber
\\
-3\eta\sum_{n}\sum\limits_{\mu q}(-1)^qeD_qC^{j\mu}_{1\bar{q},\,
j''\mu''}\frac{\langle {\rm 1s}\mu'|A^{\rm P}({\bf e},{\bf
k})|n\rangle \langle n|A^{\rm P}({\bf e'},{\bf k'})|{\rm
2p}\mu\rangle}{E_n-E_A+\omega'}+
\\
\nonumber +\sum_{n}\frac{\langle {\rm 1s}\mu'|A^{\rm P}({\bf
e'},{\bf k'})|n\rangle \langle n|A^{\rm P}({\bf e},{\bf k})|{\rm
2s}\mu''\rangle}{E_n-E_A+\omega}\hspace{1cm}
\\
\nonumber -3\eta\sum_{n}\sum\limits_{\mu
q}(-1)^qeD_qC^{j\mu}_{1\bar{q},\,j''\mu''}\frac{\langle {\rm
1s}\mu'|A^{\rm P}({\bf e'},{\bf k'})|n\rangle \langle n|A^{\rm
P}({\bf e},{\bf k})|{\rm 2p}\mu\rangle}{E_n-E_A+\omega}
 \Big|^2d{\bf n}_{\bf k}d{\bf n}_{\bf k'}d\omega',
\end{eqnarray}
where $A^{\rm P}({\bf e},{\bf k})$ is defined by
Eq.~(\ref{E1E2decomp}).

The decomposition Eq.~(\ref{E1E2decomp}) of the $A^{\rm P}({\bf
e},{\bf k})$ operator shows that the first and third terms in
Eq.~(\ref{1.extern}) correspond to the E1E1 two-photon decay rate
of the 2s electron level. Other terms in Eq.~(\ref{1.extern})
represent the admixed E1E2 and E1M1 amplitudes of the 2p state
two-photon decay probability. Application of the expansion
Eq.~(\ref{E1E2decomp}) to the second term in Eq.~(\ref{1.extern})
leads to the expression
\begin{eqnarray}\label{2.extern}
\sum_{n}\frac{\langle {\rm 1s}\mu'|A^{\rm P}({\bf e},{\bf
k})|n\rangle \langle n|A^{\rm P}({\bf e'},{\bf k'}) |{\rm
2p}\mu\rangle}{E_n-E_A+\omega'}=
-i\omega\omega'^2\sum\limits_n\frac{\langle {\rm 1s}\mu'|{\bf
e}{\bf r}|n\rangle\langle n|\frac{1}{2}({\bf e'}{\bf r})({\bf
n}_{\bf k'}{\bf r})|{\rm 2p}\mu\rangle}{E_n-E_A+\omega'}\nonumber
\\
+\omega\omega'\sum\limits_n\frac{\langle {\rm 1s}\mu'|{\bf
e}^*{\bf r}|n\rangle\langle n|\frac{1}{2}\left([{\bf e'}\times{\bf
n}_{\bf k'}](\hat{\bf l}+2\hat{\bf s})\right)|{\rm
2p}\mu\rangle}{E_n-E_A+\omega'} \hspace{1cm}
\\
\nonumber -i\omega'\omega^2\sum\limits_n\frac{\langle {\rm
1s}\mu'|\frac{1}{2}({\bf e}{\bf r})({\bf n}_{\bf k}{\bf
r})|n\rangle\langle n|{\bf e'}{\bf r}|{\rm
2p}\mu\rangle}{E_n-E_A+\omega'}\hspace{1cm}
\\
\nonumber +\omega\omega'\sum\limits_n \frac{\langle {\rm
1s}\mu'|\frac{1}{2}\left([{\bf e}\times{\bf n}_{\bf k}](\hat{\bf
l}+2\hat{\bf s})\right)|n\rangle\langle n|{\bf e'}^*{\bf r'}|{\rm
2p}\mu\rangle}{E_n-E_A+\omega'}.
\end{eqnarray}
In order to perform the summation over intermediate states in this
expression we employ the Coulomb Green function method, which was
described in the third section of this paper. We should note that
in terms involving a magnetic dipole photon the Coulomb Green
function does not occure due to the orthogonality of the radial
functions.

Summing over all angular momenta projections in the expression
(\ref{1.extern}) we obtain
\begin{eqnarray}\label{3.extern}
dW^{(2\gamma)}_{\overline{\rm 2s}\,{\rm
1s}}(\omega)=e^4\frac{\omega\omega'}{2(2\pi)^3}\left[
\frac{2}{9}\omega^2\omega'^2\left|{\bf e}{\bf
e'}\right|^2[I_1^{\rm 2s\,1s}(\omega)+ I_1^{\rm
2s\,1s}(\omega')]^2\right.\hspace{1cm} \nonumber
\\
\left. +\omega^2\omega'^2\frac{2\sqrt{3}}{3\cdot
5}\frac{\Gamma_{\rm 2p}}{\Delta^2}\left|{\bf e}{\bf
e'}\right|^2[I_1^{\rm 2s\,1s}(\omega)+ I_1^{\rm
2s\,1s}(\omega')][\omega'I_1^{\rm 2p\,1s}(\omega')(e{\bf D}{\bf
n}_{\bf k'})+\omega I_1^{\rm 2p\,1s}(\omega)(e{\bf D}{\bf n}_{\bf
k})]\hspace{1cm} \nonumber\right.
\\
\left.
+\omega\omega'2\sqrt{2}\left(\frac{2}{3}\right)^5\frac{\Gamma_{\rm
2p}}{\Delta^2}[I_1^{\rm 2s\,1s}(\omega)+I_1^{\rm
2s\,1s}(\omega')]\left[\omega\left({\bf e}\left[e{\bf
D}\times[{\bf e'}\times{\bf n}_{\bf k'}]\right]\right)+
\omega'\left({\bf e'}\left[e{\bf D}\times[{\bf e}\times{\bf
n}_{\bf k}]\right]\right)\right]({\bf e}{\bf e'})^*\hspace{1cm}
\right. \nonumber
\\
\nonumber \left. +\omega^2\omega'^2\frac{2\sqrt{3}}{3\cdot
5}\frac{\Gamma_{\rm 2p}}{\Delta^2}\left|{\bf e}{\bf
e'}\right|^2[I_1^{\rm 2s\,1s}(\omega)+ I_1^{\rm
2s\,1s}(\omega')][\omega I_2^{\rm 2p\, 1s}(\omega')(e{\bf D}{\bf
n}_{\bf k})+\omega' I_2^{\rm 2p\,1s}(\omega)(e{\bf D}{\bf n}_{\bf
k'})]\hspace{1cm} \right.
\\
\nonumber \left. +\frac{2\cdot 3}{5^2\Delta^2}\left[({\bf e}{\bf
e'})({\bf D}{\bf n}_{\bf k'}) \omega'I^{\rm
2p\,1s}_1(\omega')+({\bf e}{\bf e'})({\bf D}{\bf n}_{\bf k})\omega
I^{\rm 2p\,1s}_1(\omega)\right]^2\hspace{1cm} \right.
\\
\nonumber \left. +\frac{2\cdot 3}{5^2\Delta^2}\left[({\bf e}{\bf
e'})({\bf D}{\bf n}_{\bf k})\omega I^{\rm 2p\,1s}_2(\omega')+({\bf
e}{\bf e'})({\bf D}{\bf n}_{\bf k'})\omega'I^{\rm
2p\,1s}_2(\omega)\right]^2\hspace{1cm} \right.
\\
\nonumber \left. +\frac{4\cdot 3}{5^2\Delta^2}\left[({\bf e}{\bf
e'})({\bf D}{\bf n}_{\bf k'}) \omega'I^{\rm
2p\,1s}_1(\omega')+({\bf e}{\bf e'})({\bf D}{\bf n}_{\bf k})\omega
I^{\rm 2p\,1s}_1(\omega)\right] \left[({\bf e}{\bf e'})({\bf
D}{\bf n}_{\bf k})\omega I^{\rm 2p\,1s}_2(\omega')+({\bf e}{\bf
e'})({\bf D}{\bf n}_{\bf k'}) \omega'I^{\rm
2p\,1s}_2(\omega)\right] \right.
\\
\nonumber \left. + \frac{2^{14}}{3^7\Delta^2}\left[
\omega\left([{\bf D}\times{\bf e}][{\bf n}_{\bf k'}\times{\bf
e'}]\right) + \omega'\left([{\bf D}\times{\bf e'}][{\bf n}_{\bf
k}\times{\bf e}]\right)\right]^2 \hspace{1cm} \right.
\\
\left. +\frac{2^{15}}{3^7\Delta^2} \left[ \omega\left({\bf D}{\bf
e}\right) [{\bf n}_{\bf k'}\times{\bf e'}] +\omega'\left({\bf
D}{\bf e'}\right)[{\bf n}_{\bf k}\times{\bf e}]\right]^2
\right]d{\bf n}_{\bf k}d{\bf n}_{\bf k'} d\omega',
\end{eqnarray}
where $I_1^{\rm 2s\,1s}(\omega)$ is defined by
Eq.~(\ref{2s.rad.1}) and Eq.~(\ref{2s.rad.2}) for the 2s-1s
transition, and $I_1^{\rm 2p\,1s}(\omega)$, $I_2^{\rm
2p\,1s}(\omega)$ are defined by Eqs.~(\ref{4.5}), (\ref{4.6}),
correspondingly.

The summation over polarizations is provided by Eqs.~(\ref{1.Fik})
and (\ref{E1E2.pol}). The radial integration can be analytically
performed as well. Finally, the integration over $\omega'$ is
carried out with MATHEMATICA code. The result is
\begin{eqnarray}\label{4.extern}
\frac{dW^{(2\gamma)}_{\rm \overline{2s}\,1s}}{d{\bf n}_{\bf
k}d{\bf n}_{\bf k'}}=0.00131822(\alpha Z)^6
-\frac{0.000230135}{\pi^3}\frac{e\Gamma_{\rm 2p}}{\Delta^2}[{\bf
D}{\bf n}_{\bf k}+{\bf D}{\bf n}_{\bf k'}]\left(1+({\bf n}_{\bf
k}{\bf n}_{\bf k'})^2\right)(\alpha Z)^7\nonumber
\\
-\frac{0.0000340919}{\pi^3}\frac{e\Gamma_{\rm 2p}}{\Delta^2}[{\bf
D}{\bf n}_{\bf k}+{\bf D}{\bf n}_{\bf k'}]\left(1+({\bf n}_{\bf
k}{\bf n}_{\bf k'})\right)(\alpha
Z)^7+0.00175091\frac{e^2D^2}{\Delta^2}(\alpha Z)^8.
\end{eqnarray}
The first term in this expression is the differential E1E1 2s-1s
transition probability. The second and third terms represent
interference terms for the mixed 2s and 2p two-photon transition
probabilities (second term corresponds to the E1E1$\cdot$E1E2
transition probability and third one represents E1E1$\cdot$E1M1
transition probability). Last term is related to the sum of the
E1E2 and E1M1 two-photon probabilities, respectively.

Rewriting Eq.~(\ref{4.extern}) in a form similar to Eq.~(\ref{1.19}) yields
\begin{eqnarray}\label{5.extern}
\frac{dW^{(2\gamma)}_{\rm\overline{2s}\,1s}}{d{\bf n}_{\bf k}d{\bf
n}_{\bf k'}} = W_0\left[1\pm \beta_1(D)\left[{\bf n}_{\bf D}{\bf
n}_{\bf k}+{\bf n}_{\bf D}{\bf n}_{\bf k'}\right](1+({\bf n}_{\bf
k}{\bf n}_{\bf k'})^2)\pm \beta_2(D)\left[{\bf n}_{\bf D}{\bf
n}_{\bf k}+{\bf n}_{\bf D}{\bf n}_{\bf k'}\right](1+({\bf n}_{\bf
k}{\bf n}_{\bf k'}))\right],
\end{eqnarray}
where $W_0=W_{\rm 2s}^{(2\gamma)}+\widetilde{W}_{\rm
2p}^{(2\gamma)}e^2D^2/\Delta^2$, $\widetilde{W}_{\rm
2p}^{(2\gamma)}$ is the sum of the E1E2 and E1M1 transition
probabilities.

Functions $\beta_1(D)$ and $\beta_2(D)$ are defined by
\begin{eqnarray}\label{5.beta}
\beta_1(D)=\frac{0.000230135(\alpha
Z)^7}{W_0\pi^3}\frac{|e|D\Gamma_{\rm 2p}}{\Delta^2},
\\
\nonumber \beta_2(D)=\frac{0.0000340919(\alpha
Z)^7}{W_0\pi^3}\frac{|e|D\Gamma_{\rm 2p}}{\Delta^2}.
\end{eqnarray}
The maximum value for $\beta_1$ (or $\beta_2$) is achieved at the
field strength
\begin{eqnarray}\label{5.maximums}
D_{\rm max}=\frac{w^{2\gamma}\Delta}{|e|}\approx \pm 0.000018\,
{\rm a.u.} \approx \pm 57\, {\rm kV/cm},
\end{eqnarray}
where the $(-)$ and $(+)$ signs in Eq.~(\ref{dmax}) correspond to
the H and $\overline{{\rm H}}$ atoms, respectively.

Then the corresponding maximum value of the transition rate
$dW^{(2\gamma)}_{\rm \overline{2s}\,1s}$ is obtained via
\begin{eqnarray}\label{dmax}
\frac{dW^{(2\gamma)}_{\rm\overline{2s}\,1s}}{d{\bf n}_{\bf k}d{\bf
n}_{\bf k'}} = W_0(D_{\rm max})\qquad
\nonumber
\\
\times\left[1\pm 0.00024397\left[{\bf
n}_{\bf D}{\bf n}_{\bf k}+{\bf n}_{\bf D}{\bf n}_{\bf
k'}\right](1+({\bf n}_{\bf k}{\bf n}_{\bf k'})^2)\pm
0.00003614\left[{\bf n}_{\bf D}{\bf n}_{\bf k}+{\bf n}_{\bf D}{\bf
n}_{\bf k'}\right](1+({\bf n}_{\bf k}{\bf n}_{\bf k'}))\right].
\end{eqnarray}
Integration over the directions of the emitted photons ${\bf
n}_{\bf k}$ and ${\bf n}_{\bf k'}$ leads to the value
Eq.~(\ref{3.58}) for the E1E1 two-photon decay rate of the
2s-state, because the interference terms give zero result (see
Eq.~(\ref{E1E2.6})). But for the differential transition
probability Eq.~(\ref{5.extern}) the interference terms clearly
demonstrate the linear dependence on the external electric field
${\bf D}$. Thus the total two-photon $W^{(2\gamma)}_{\rm
\overline{2s}\,1s}$ transition probability integrated over photon
directions is:
\begin{eqnarray}\label{trpr}
W^{(2\gamma)}_{\rm\overline{2s}\,1s}(D_{\rm max})=W_0(D_{\rm
max})=W_{\rm 2s}^{(2\gamma)}+\frac{\widetilde{W}_{\rm
2p}^{(2\gamma)}e^2D_{\rm max}^2}{\Delta^2}\approx3.98116\cdot
10^{-16}\, {\rm a.u.}\approx16.4585\, {\rm s}^{-1},
\end{eqnarray}
i.e. twice as large as the zero field value Eq.~(\ref{2E1zf}).

In principle, the dependece on the external electric field in the
transition probability $W^{(2\gamma)}_{\rm \overline{2s}\,1s}$
Eq.~(\ref{trpr}) can be considered as a correction which does not
vanish after integration over the photons emission directions. If
we return to the  radiative correction considered in
\cite{Jentschura}, then it is easy to see that the radiative
correction (Eq.~(36) in \cite{Jentschura})
$\delta\Gamma_{2s}/\Gamma_{2s}=−
2.020536\frac{\alpha}{\pi}(\alpha Z)^{2}\ln\left[(\alpha
Z)^{-2}\right]=-2.4594\cdot10^{-6}$ corresponds to the magnitude
of the field $|D_{\rm r}|\approx D_{\rm
max}\sqrt{\delta\Gamma_{\rm 2s}/\Gamma_{\rm 2s}}\approx 2.8\cdot
10^{-8}\,{\rm a.u.}\approx 90$ V/cm. Such fields are often used in
the spectroscopic experiments, therefore, this effect also should
be included in this context.

The linear over field corrections $\beta_1(D)$ and $\beta_2(D)$ in
Eq.~(\ref{5.extern}) reach the magnitude of the radiative
correction at the fields approximately of the same order. Unlike
the correction, discussed earlier in the section ``2s decay rate
for hydrogen and anti-hydrogen atoms in external electric
fields'', this is the correction directly to the same process, as
radiative correction \cite{Jentschura}.

As it was mentioned above the formal T-noninvariance of the factor
${\bf n}_{\bf D}{\bf n}_{\bf k}$ in Eq.~(\ref{1.dW}) and in
Eq.~(\ref{5.extern}) (${\bf n}_{\bf k}$ and ${\bf n}_{\bf D}$ are
T-odd and T-even vectors, respectively) is compensated by the
dependence on $\Gamma_{\rm 2p}$; This is the imitation of
T-noninvariance  in unstable systems, as predicted by Zeldovich
\cite{Zeldovich}.

The relative difference for the decay rates in H and
$\overline{{\rm H}}$ atoms at the maximum value $D_{\rm max}$
equals to:
\begin{eqnarray}
\frac{dW_{\rm \overline{2s}\, 1s}^{(2\gamma)}(\rm H)}{W_0(D_{\rm
max})d{\bf n}_{\bf k}d{\bf n}_{\bf k'}}-\frac{dW_{\rm
\overline{2s}\, 1s}^{(2\gamma)}(\overline{{\rm H}})}{W_0(D_{\rm
max})d{\bf n}_{\bf k}d{\bf n}_{\bf k'}}=\qquad \\
\nonumber = 2\beta_1(D_{\rm max})({\bf n}_{\bf D}{\bf n}_{\bf k}+{\bf n}_{\bf D}{\bf n}_{\bf k'})(1+({\bf n}_{\bf k}{\bf n}_{\bf k'})^2)+2\beta_2(D_{\rm max})({\bf n}_{\bf D}{\bf n}_{\bf k}+{\bf n}_{\bf D}{\bf n}_{\bf k'})(1+{\bf n}_{\bf k}{\bf n}_{\bf k'})  \\
\nonumber = ({\bf n}_{\bf D}{\bf n}_{\bf k}+{\bf n}_{\bf D}{\bf
n}_{\bf k'})(0.000280111+0.00024397({\bf n}_{\bf k}{\bf n}_{\bf
k'})+0.0000361414{\bf n}_{\bf k}{\bf n}_{\bf k'}).
\end{eqnarray}
This ratio is close to 0.028\% and
represents a tiny effect reflecting the difference between matter
and anti-matter even at maximum field strength $D_{{\rm max}}$.

For completeness the $\overline{\rm 2p}\to {\rm 1s}+2\gamma$
two-photon transition probability should be considered as well. It
can be evaluated similarly to $\overline{\rm 2s}\to {\rm
1s}+2\gamma$ two-photon decay rate with the use of the wave
function
\begin{eqnarray}\label{6.extern}
|\overline{\rm 2p}\mu''\rangle =|{\rm 2p}
 \mu''\rangle - \eta\sum\limits_{\mu}
\langle {\rm 2s}\mu''|e{\bf D}{\bf r}|{\rm 2p}\mu\rangle|{\rm
2s}\mu''\rangle.
\end{eqnarray}
In this case the two-photon transition without an external
electric field will be provided by the sum of the E1E2 and E1M1
decays, and the interference terms will be the same as in
Eq.~(\ref{3.extern}).

The result can be presented in the form
\begin{eqnarray}\label{5.extern.1}
dW_{\rm \overline{2p}\,1s}^{(2\gamma)}=\left[dW_{\rm 2p\,1s}^{\rm
E1E2}+dW_{\rm 2p\,1s}^{\rm E1M1}+\frac{9e^2D^2}{\Delta^2}dW_{\rm
2s\,1s}^{\rm E1E1}+\frac{0.000230135}{\pi^3}\frac{\Gamma_{\rm
2p}}{\Delta^2}[e{\bf D}{\bf n}_{\bf k}+e{\bf D}{\bf n}_{\bf
k'}]\left(1+({\bf n}_{\bf k}{\bf n}_{\bf k'})^2\right)(\alpha Z)^7
\nonumber\right.
\\
\left. +\frac{0.0000340919}{\pi^3}\frac{\Gamma_{\rm
2p}}{\Delta^2}[e{\bf D}{\bf n}_{\bf k}+e{\bf D}{\bf n}_{\bf k'}]
\left(1+{\bf n}_{\bf k}{\bf n}_{\bf k'}\right)(\alpha
Z)^7\right]\,d{\bf n}_{\bf k}d{\bf n}_{\bf k'},
\end{eqnarray}
\begin{eqnarray}\label{5.extern.2}
\frac{dW^{(2\gamma)}_{\rm \overline{2p}\, 1s}}{d{\bf n}_{\bf
k}d{\bf n}_{\bf k'}} = W_0\left[1\mp \beta_1(D)\left[{\bf n}_{\bf
D}{\bf n}_{\bf k}+{\bf n}_{\bf D}{\bf n}_{\bf k'}\right](1+({\bf
n}_{\bf k}{\bf n}_{\bf k'})^2) \mp \beta_2(D)\left[{\bf n}_{\bf
D}{\bf n}_{\bf k}+{\bf n}_{\bf D}{\bf n}_{\bf k'}\right](1+{\bf
n}_{\bf k}{\bf n}_{\bf k'})\right],
\end{eqnarray}
where $W_0=W^{\rm (E1E2)}_{\rm 2p\,1s}+W^{\rm (E1M1)}_{\rm 2p\,
1s}+9e^2D^2W^{\rm (E1E1)}_{\rm 2s\,1s}/\Delta^2$ and the functions
$\beta_1(D)$, $\beta_2(D)$ are defined again by
Eq.~(\ref{5.beta}).

Then the maximum of the $\beta_1$ (or $\beta_2$) is achieved at
\begin{eqnarray}\label{5.maximums.2}
|D_{{\rm max}}| =\frac{\Delta}{3|e|w^{2\gamma}}\approx 7.1\cdot
10^{-11}\, {\rm a.u.}\approx 0.23\, {\rm V/cm}.
\end{eqnarray}
The corresponding maximum value of $dW^{(2\gamma)}_{\rm
\overline{2p}\,1s}$ is
\begin{eqnarray}\label{dmax.2}
\frac{dW^{(2\gamma)}_{\rm \overline{2p}\,1s}}{d{\bf n}_{\bf
k}d{\bf n}_{\bf k'}}=W_0(D_{{\rm max}})\qquad
\nonumber
\\
\times
\left[1\mp
0.00048613\left[{\bf n}_{\bf D}{\bf n}_{\bf k}+{\bf n}_{\bf D}{\bf
n}_{\bf k'} \right](1+({\bf n}_{\bf k}{\bf n}_{\bf k'})^2)\mp
0.000720147\left[{\bf n}_{\bf D}{\bf n}_{\bf k}+ {\bf n}_{\bf
D}{\bf n}_{\bf k'}\right](1+{\bf n}_{\bf k}{\bf n}_{\bf
k'})\right].
\end{eqnarray}
After integration over ${\bf n}_{\bf k}$ and ${\bf n}_{\bf k'}$ in
Eq.~(\ref{5.extern.1}) the term quadratic in the external electric
field for the correction to the two-photon transition probability
2p-1s still remains:
\begin{eqnarray}
W^{(2\gamma)}_{\rm \overline{2p}\,1s}(D_{\rm max}) = W^{\rm
(E1E2)}_{\rm 2p\, 1s}+W^{\rm (E1M1)}_{\rm 2p\,
1s}+\frac{9e^2D_{\rm max}^2W^{\rm (E1E1)}_{\rm 2s\, 1s}}{\Delta^2}
\approx 4.09\cdot 10^{-22}\, {\rm a.u.}\approx 1.69\cdot 10^{-5}\,
{\rm s}^{-1}.
\end{eqnarray}
Finally, the $(+)$ and $(-)$ signs in Eq.~(\ref{dmax.2})
correspond to the  H  and $\overline{{\rm H}}$ atoms,
respectively. The relative difference for the decay rates in H and
$\overline{{\rm H}}$ atoms at the maximum value $D_{\rm max}$
equals to:
\begin{eqnarray}\label{1.relative}
\frac{dW_{\rm \overline{2p}\, 1s}^{(2\gamma)}}{W_0(D_{{\rm
max}})d{\bf n}_{\bf k}d{\bf n}_{\bf k'}}({\rm H})-\frac{dW_{\rm
\overline{2p}\,
1s}^{(2\gamma)}}{W_0(D_{{\rm max}})d{\bf n}_{\bf k}d{\bf n}_{\bf k'}}(\overline{{\rm H}})=\qquad \\
\nonumber = 2\beta_1(D_{{\rm max}}) ({\bf n}_{\bf D}{\bf n}_{\bf k}+{\bf n}_{\bf D}{\bf n}_{\bf k'})(1+({\bf n}_{\bf k}{\bf n}_{\bf k'})^2)+2\beta_2(D_{\rm max}) ({\bf n}_{\bf D}{\bf n}_{\bf k}+{\bf n}_{\bf D}{\bf n}_{\bf k'})(1+{\bf n}_{\bf k}{\bf n}_{\bf k'}) \\
\nonumber = ({\bf n}_{\bf D}{\bf n}_{\bf k})+{\bf n}_{\bf D}{\bf
n}_{\bf k'})(0.0111629 + 0.0097226({\bf n}_{\bf k}{\bf n}_{\bf
k'})^2 + 0.00144029{\bf n}_{\bf k}{\bf n}_{\bf k'}).
\end{eqnarray}
This ratio turns out to be close to $1$\%. However, any direct
observation of this difference should be difficult due to the huge
background from the one-photon transition ${\rm 2p}\rightarrow
{\rm 1s}+\gamma$.

\section{Three-photon decay rate for the $2p$ state of hydrogen-like light atomic systems.}

In this last section we present the calculation of the E1E1E1
transition probability for the 2p hydrogenic state decay.
Parametric estimate can be easily obtained in usual way and it is
$\alpha(\alpha Z)^8$ a.u. Therefore, one can expect a numerical
result for the E1E1E1 decay rate which is comparable with values
obtained for the E1E2 and E1M1 transition.

According to the Feynman rules the $S$-matrix element for the
$n$-photons emission process with the transition from the state
$A$ to $A'$ ($A\rightarrow n\gamma + A'$) can be written as (in
r.u.):
\begin{eqnarray}
S_{A'A}^{(n)}=(-ie)^n\int\left(\bar{\psi}_{A'}(x_1)\hat{A}(x_1)S(x_1,x_2)\hat{A}(x_2)
...\hat{A}(x_{n-1})S(x_{n-1},x_n)\psi_A(x_n)\right)dx_1...dx_n
 \label{S-matrix}
\end{eqnarray}
where $\hat{A}$ is the emission operator, $\psi_A, \overline\psi_{A'}$ are
Dirac wave function, $S(x_{n-1},x_n)$ etc. denotes the bound electron
propagator, and $e$ is the electron charge.

The emission operator is given by
\begin{eqnarray} \label{emOper}
\hat{A}(x)=\sqrt{\frac{2\pi}{\omega}}\hat{e}^{(\lambda)}e^{-i({\bf
k}{\bf r}-\omega t)},
\end{eqnarray}
where $\hat{e}^{(\lambda)}$ is the 4-vector of the photon
polarization ($\hat{e}^{(\lambda)}={\bf e}\gamma$, ${\bf e}$ is
the polarization vector, $\gamma$ are the Dirac matrices) and ${\bf
k}$ is the wave vector. In what follows we employ the
nonrelativistic approximation, replacing the Dirac solutions for
atomic electron orbitals by the corresponding Schr\"{o}dinger ones
and omitting, where it is justified, the contributions of the
negative energy states to the exact electron propagators in
Eq.~(\ref{S-matrix}).

Integrating over time variables in Eq.~(\ref{S-matrix}) results in
\begin{eqnarray} \label{S-matr}
S_{A'A}^{(n)}=-2\pi
i(-ie)^n\left(\frac{2\pi}{\omega}\right)^{n/2}\delta(E_{A'}+n\omega-E_{A})\sum\limits_{s_1,...,s_{n-1}}
\frac{\left(\hat{e}e^{i{\bf k}{\bf
r}}\right)_{A's_{n-1}}...\left(\hat{e}e^{i{\bf k}{\bf
r}}\right)_{s_{1}A}}
{\left[E_{s_{n-1}}-E_A-(n-1)\omega\right]...\left[E_{s_1}-E_A-\omega\right]},
\end{eqnarray}
where the summation over $s_i$ extends over all
intermediate Schr\"odinger states with positive energy, $E_s$ are the
Schr\"{o}dinger energies for an atomic electron.

In the nonrelativistic approximation we can expand the exponents
$e^{i{\bf k}{\bf r}}$ in Eq.~(\ref{S-matr}), leaving only the
first term and replace the matrix elements $(\hat{e})_{ss'}$ with
Dirac wave functions by the matrix elements $({\bf e}\hat{\bf
p})_{ss'}$, with Schr\"{o}dinger wave functions, where
$\hat{\hat{\bf p}}$ is the electron momentum operator. Then, using
a well known quantum mechanical relation
 $\left([{\bf r},\hat{H}]_-\right)_{ss'}=i(\hat{\bf p})_{ss'}$, the
amplitude of the three-photon emission process can be written in a
form:
\begin{eqnarray}\label{S3}
U_{A'A}^{(3)}=-(-ie)^3(2\pi)^{3/2}\sqrt{\omega_1\omega_2\omega_3}\,\delta(E_{A'}+\omega_1+\omega_2+
\omega_3-E_{A})\times \nonumber
\\
\left[\sum\limits_{s_1,s_{2}} \frac{\left({{\bf e}^*_1}{{\bf
r}_1}\right)_{A's_{2}}\left({{\bf e}^*_2}{{\bf
r}_2}\right)_{s_{2}s_1}\left({{\bf e}^*_3}{{\bf
r}_3}\right)_{s_{1}A}}
{(E_{s_2}-E_A+\omega_1+\omega_2)(E_{s_1}-E_A+\omega_3)}+
\sum\limits_{s_1,s_{2}} \frac{\left({{\bf e}^*_1}{{\bf
r}_1}\right)_{A's_{2}}\left({{\bf e}^*_3}{{\bf
r}_3}\right)_{s_{2}s_1}\left({{\bf e}^*_2}{{\bf
r}_2}\right)_{s_{1}A}}
{(E_{s_2}-E_A+\omega_1+\omega_3)(E_{s_1}-E_A+\omega_2)}\right.
\nonumber
\\
\left. +\sum\limits_{s_1,s_{2}} \frac{\left({{\bf e}^*_2}{{\bf
r}_2}\right)_{A's_{2}}\left({{\bf e}^*_1}{{\bf
r}_1}\right)_{s_{2}s_1}\left({{\bf e}^*_3}{{\bf
r}_3}\right)_{s_{1}A}}
{(E_{s_2}-E_A+\omega_1+\omega_2)(E_{s_1}-E_A+\omega_3)} +
\sum\limits_{s_1,s_{2}} \frac{\left({{\bf e}^*_2}{{\bf
r}_2}\right)_{A's_{2}}\left({{\bf e}^*_3}{{\bf
r}_3}\right)_{s_{2}s_1}\left({{\bf e}^*_1}{{\bf
r}_1}\right)_{s_{1}A}}
{(E_{s_2}-E_A+\omega_2+\omega_3)(E_{s_1}-E_A+\omega_1)}\right.
\\
\nonumber \left. +\sum\limits_{s_1,s_{2}} \frac{\left({{\bf
e}^*_3}{{\bf r}_3}\right)_{A's_{2}}\left({{\bf e}^*_1}{{\bf
r}_1}\right)_{s_{2}s_1}\left({{\bf e}^*_2}{{\bf
r}_2}\right)_{s_{1}A}}
{(E_{s_2}-E_A+\omega_1+\omega_3)(E_{s_1}-E_A+\omega_2)} +
\sum\limits_{s_1,s_{2}} \frac{\left({{\bf e}^*_3}{{\bf
r}_3}\right)_{A's_{2}}\left({{\bf e}^*_2}{{\bf
r}_2}\right)_{s_{2}s_1}\left({{\bf e}^*_1}{{\bf
r}_1}\right)_{s_{1}A}}
{(E_{s_2}-E_A+\omega_2+\omega_3)(E_{s_1}-E_A+\omega_1)}\right].
\end{eqnarray}
Here $\omega_1$, $\omega_2$, $\omega_3$ are the frequencies (energies) of the emitted photon.

 The probability of the three-photon emission process is
\begin{eqnarray}\label{prob1}
dW^{(3)}_{A'A}=2\pi
\left|U_{A'A}^{(3)}\right|^2\delta(E_{A'}+\omega_1+\omega_2+\omega_3-E_{A})\frac{d{\bf
k_1}d{\bf k_2}d{\bf k_3}}{(2\pi)^9}.
\end{eqnarray}
The total probability can be received from Eq.~(\ref{prob1}) by
the summation over photon polarizations ${\bf e}_1, {\bf e}_2,
{\bf e}_3$ and integration over all the photon-emission directions
${\bf k}_1, {\bf k}_2, {\bf k}_3$ and frequencies $\omega_1,
\omega_2, \omega_3$. For the summation over polarizations and
integration over photon directions it is convenient to use the
relations Eqs.~(\ref{1.Fik}), (\ref{E1E2.6}), (\ref{E1E2.pol}).

After averaging over angular momentum projections of the initial state
and summing over final ones, the probability of the three-photon emission process
results as
\begin{eqnarray} \label{9}
dW_{AA'}^{(3)}=e^92^4\omega_1^3\omega_2^3\omega_3^3\frac{d\omega_1d\omega_2d\omega_3}{\pi^2}
\delta(E_{A'}+\omega_1+\omega_2+\omega_3-E_{A})\times
\\
\nonumber \sum\limits_{m_{l_{A'}}m_{l_A}}\frac{1}{2l_{A}+1}
\sum_{q_1q_2q_3}(-1)^{q_1+q_2+q_3}U_{A'A}^{(3)}(q_1,q_2,q_3)U_{A'A}^{(3)*}(-q_1,-q_2,-q_3)\, ,
\end{eqnarray}
where the expression for the amplitude $U_{A'A}^{(3)}(q_1,q_2,q_3)$ in spherical representation
is given by
\begin{eqnarray}\label{components}
U_{A'A}^{(3)}(q_1,q_2,q_3)=\left[
\sum\limits_{s_1s_2}\frac{\left(({\bf
r}_1)_{q_1}\right)_{A's_2}\left(({\bf r}_2)_{q_2}\right)_{s_2s_1}
\left(({\bf
r}_3)_{q_3}\right)_{s_1A}}{(E_{s_2}-E_A+\omega_1+\omega_2)(E_{s_1}-E_A+\omega_3)}+
\sum\limits_{s_1s_2}\frac{\left(({\bf
r}_1)_{q_1}\right)_{A's_2}\left(({\bf r}_3)_{q_3}\right)_{s_2s_1}
\left(({\bf
r}_2)_{q_2}\right)_{s_1A}}{(E_{s_2}-E_A+\omega_1+\omega_3)(E_{s_1}-E_A+\omega_2)}
\nonumber \right.
 \\
 \left.
+\sum\limits_{s_1s_2}\frac{\left(({\bf
r}_2)_{q_2}\right)_{A's_2}\left(({\bf r}_1)_{q_1}\right)_{s_2s_1}
\left(({\bf
r}_3)_{q_3}\right)_{s_1A}}{(E_{s_2}-E_A+\omega_1+\omega_2)(E_{s_1}-E_A+\omega_3)}+
\sum\limits_{s_1s_2}\frac{\left(({\bf
r}_2)_{q_2}\right)_{A's_2}\left(({\bf r}_3)_{q_3}\right)_{s_2s_1}
\left(({\bf
r}_1)_{q_1}\right)_{s_1A}}{(E_{s_2}-E_A+\omega_2+\omega_3)(E_{s_1}-E_A+\omega_1)}
 \right. \label{amplitude}
 \\
 \left. \nonumber
+\sum\limits_{s_1s_2}\frac{\left(({\bf r}_3)_{q_3}\right)_{A's_2}\left(({\bf r}_1)_{q_1}\right)_{s_2s_1}
\left(({\bf r}_2)_{q_2}\right)_{s_1A}}{(E_{s_2}-E_A+\omega_1+\omega_3)(E_{s_1}-E_A+\omega_2)}+
\sum\limits_{s_1s_2}\frac{\left(({\bf r}_3)_{q_3}\right)_{A's_2}\left(({\bf r}_2)_{q_2}\right)_{s_2s_1}
\left(({\bf r}_1)_{q_1}\right)_{s_1A}}{(E_{s_2}-E_A+\omega_2+\omega_3)(E_{s_1}-E_A+\omega_1)}
 \right].
\end{eqnarray}
In order to calculate the transition probabilities for the process
${\rm 2p}\rightarrow 3\gamma({\rm E1})+{\rm 1s}$ in the hydrogen
atom the nonrelativistic Coulomb Green function method is
employed. Inserting the Green function in a form Eq.~(\ref{3.22})
in (\ref{amplitude}) and representing the vector component $({\bf
r})_q$ like $r_q=\sqrt{\frac{4\pi}{3}}Y_{1q}$ we can perform the
angular integration, which gives
\begin{eqnarray}\label{AngInt}
U^{(3)}_{A'A}(q_1,q_2,q_3)=\sqrt{\frac{2l_A+1}{2l_{A'}+1}}\sum\limits_{l_1m_{l_1}}\sum\limits_{l_2m_{l_2}}
C_{10\, l_20}^{l_{A'}0}C_{10\, l_10}^{l_20}C_{10\,
l_A0}^{l_10}\big\{F_{l_1l_2}(\nu_1,\nu_2)\left[C_{1q_1\,
l_2m_{l_2}}^{l_{A'}m_{A'}}C_{1q_2\,
l_1m_{l_1}}^{l_2m_{l_2}}C_{1q_3\, l_Am_A}^{l_1m_{l_1}}
 \nonumber \right.
\\ \left.
C_{1q_2\, l_2m_{l_2}}^{l_{A'}m_{A'}}C_{1q_1\,
l_1m_{l_1}}^{l_2m_{l_2}}C_{1q_3\,
l_Am_A}^{l_1m_{l_1}}\right]+F_{l_1l_2}(\nu_3,\nu_4)\left[C_{1q_1\,
l_2m_{l_2}}^{l_{A'}m_{A'}}C_{1q_3\,
l_1m_{l_1}}^{l_2m_{l_2}}C_{1q_2\, l_Am_A}^{l_1m_{l_1}}+ C_{1q_3\,
l_2m_{l_2}}^{l_{A'}m_{A'}}C_{1q_1\,
l_1m_{l_1}}^{l_2m_{l_2}}C_{1q_2\, l_Am_A}^{l_1m_{l_1}}\right]
\\ \nonumber
+F_{l_1l_2}(\nu_5,\nu_6)\left[C_{1q_2\,
l_2m_{l_2}}^{l_{A'}m_{A'}}C_{1q_3\,
l_1m_{l_1}}^{l_2m_{l_2}}C_{1q_1\, l_Am_A}^{l_1m_{l_1}}+ C_{1q_3\,
l_2m_{l_2}}^{l_{A'}m_{A'}}C_{1q_2\,
l_1m_{l_1}}^{l_2m_{l_2}}C_{1q_1\, l_Am_A}^{l_1m_{l_1}}\right]
\big\}.
\end{eqnarray}
Here $C_{l_1m_{l_1}\, l_2m_{l_2}}^{l_3m_{l_3}}$ is the
Clebsch-Gordan coefficient, and
\begin{eqnarray}\label{radial}
F_{l_1,
l_2}(\nu_i,\nu_j)=\int\limits_0^{\infty}\int\limits_0^{\infty}\int\limits_0^{\infty}dr_1dr_2dr_3r_1^3r_2^3r_3^3R_{n_{A'}l_{A'}}(r_1)
g_{l_2}(\nu_i;r_1,r_2)g_{l_1}(\nu_j;r_2,r_3)R_{n_A,l_A}(r_3),
\end{eqnarray}
\begin{eqnarray}\label{nu}
\nu_1=Z/\sqrt{-2(E_A-\omega_1-\omega_2)},
\nu_2=Z/\sqrt{-2(E_A-\omega_3)},
\nu_3=Z/\sqrt{-2(E_A-\omega_1-\omega_3)}
\\
\nonumber \nu_4=Z/\sqrt{-2(E_A-\omega_2)},
\nu_5=Z/\sqrt{-2(E_A-\omega_2-\omega_3)},
\nu_6=Z/\sqrt{-2(E_A-\omega_1)}
\end{eqnarray}
corresponding to the 6 different terms in Eq.~(\ref{components}).

For the ${\rm 2p}\rightarrow 3\gamma({\rm E1})+{\rm 1s}$ process,
$R_{n_A,l_A}\equiv R_{\rm 2p}$, $R_{n_{A'}l_{A'}}\equiv R_{\rm
1s}$, $l_A=1$, $l_{A'}=0$. Performing the summation over all
angular momentum projections in Eq.~(\ref{9}) we arrive at the
expression
\begin{eqnarray}\label{radial2}
dW^{(3)}_{\rm
2p\,1s}=e^9\omega_1^3\omega_2^3\omega_3^3\frac{d\omega_1d\omega_2d\omega_3}{5\cdot
3^3\pi^2}\frac{2^6}{3} \delta(E_{\rm
1s}+\omega_1+\omega_2+\omega_3-E_{\rm 2p})
\left[15F_{01}(\nu_1,\nu_2)^2+ \hspace{1.5cm}\right. \nonumber
\\ \left.
+15F_{01}(\nu_3,\nu_4)^2+15F_{01}(\nu_5,\nu_6)^2+20F_{01}(\nu_5,\nu_6)F_{21}(\nu_1,\nu_2)+12F_{21}(\nu_1,\nu_2)^2+
20F_{01}(\nu_5,\nu_6)F_{21}(\nu_3,\nu_4)
 \right.
\nonumber
\\ \left.
+4F_{21}(\nu_1,\nu_2)F_{21}(\nu_3,\nu_4)+12F_{21}(\nu_3,\nu_4)^2
+4F_{21}(\nu_1,\nu_2)F_{21}(\nu_5,\nu_6)+4F_{21}(\nu_3,\nu_4)F_{21}(\nu_5,\nu_6)\hspace{1cm}
 \right.
\nonumber
\\ \left.
+12F_{21}(\nu_5,\nu_6)^2+10F_{01}(\nu_3,\nu_4)\{F_{01}(\nu_5,\nu_6)+2F_{21}(\nu_1,\nu_2)+2F_{21}(\nu_5,\nu_6)\}\hspace{3cm}
\right.
\\
\nonumber \left.
\hspace{4cm}+10F_{01}(\nu_1,\nu_2)\{F_{01}(\nu_3,\nu_4)+F_{01}(\nu_5,\nu_6)+2F_{21}(\nu_3,\nu_4)+2F_{21}(\nu_5,\nu_6)\}
 \right].
\end{eqnarray}
After performing the radial integrations we should integrate over
frequencies. The $\delta$-function in Eq.~(\ref{radial2})
annihilates one of the integrations. It is convenient to perform
the other integrations after the transformation of the variables:
$\omega_1=\frac{1}{4}\Delta E(1-x)(1+y)$,
$\omega_2=\frac{1}{4}\Delta E(1-x)(1-y)$ and
$\omega_3=\frac{1}{2}\Delta E(1+x)$. Here $\Delta E=E_{\rm
2p}-E_{\rm 1s}=3/8$ in atomic units. The final result will be
written in atomic units:
\begin{eqnarray}\label{dW}
 W^{(3)}_{\rm 2p\,1s}=\int\limits_{0}^{3/8}d\omega_3\int\limits_0^{3/8-\omega_3}d\omega_2\int\limits_0^{3/8-\omega_3-\omega_2}d\omega_1 dW_{\rm 2p\,1s}^{(3)}
 =\frac{1}{3!}\int\limits_{0}^{3/8}\int\limits_0^{3/8}\int\limits_0^{3/8}dW_{\rm 2p\,1s}^{(3)}=
 \nonumber
 \\
 =\frac{1}{3!}\left(\frac{3}{32}\right)^2\int\limits_{-1}^{1}\int\limits_{-1}^{1}dW_{\rm 2p\,1s}^{(3)}\frac{dxdy}{(1+y)(1-x)}=
 0.263466(1)\cdot 10^{-4}\alpha(\alpha Z)^8 {\rm a.u.}=6.391(1)\cdot 10^{-8}\, {\rm s}^{-1}.
\end{eqnarray}
For $Z=1$ the result is $W_{\rm 2p\,1s}^{(3)}=6.391(1)\cdot
10^{-8}\, {\rm s}^{-1}$. The smallness of this result even
compared to the two-photon 2p-1s decays means that the 3-photon
transitions cannot play any significant role for the astrophysical
purposes, i.e. for the electron recombination history in the early
Universe. For comparison we also provide the one-, two- and
three-photon transition rates for transitions from 2p, 2s levels
(in r.u. and s$^{-1}$, where $m$ is the electron mass):
\newpage
\begin{table}
\caption{Electric (E1) and magnetic (M1) multi-photon transition
rates between 2s, 2p and 1s levels are presented for hydrogen-like
ions in relativistic units (r.u.) and for hydrogen in units
s$^{-1}$, respectively. $m$ and $Z$ denote the electron mass and
nuclear charge number.}
\begin{tabular}{ l c  r r}\hline
transition & (type) & \hspace{1cm}  transition rates \phantom{:
H-like ions} &  \hspace{2cm} \phantom{hydrogen} \\ \hline ${\rm
2p}\rightarrow {\rm 1s}$ & (E1) & $3.902\cdot 10^{-2}\,
m\alpha(\alpha Z)^4$ \, r.u.&
$6.25\cdot 10^{8}\,{\rm s}^{-1}$ \\
%
${\rm 2s}\rightarrow {\rm 1s}$ & (E1E1) &$1.317\cdot 10^{-3}\, m\alpha^2(\alpha Z)^6$ r.u.& $8.229\, {\rm s}^{-1}$\\
%
${\rm 2p}\rightarrow {\rm 1s}$& (E1E1E1) & $2.635\cdot 10^{-5}\, m\alpha^3(\alpha Z)^8$ r.u.& $6.39\cdot 10^{-8}\, {\rm s}^{-1}$ \\
${\rm 2s}\rightarrow {\rm 1s}$ & (M1) & $\frac{1}{972}\, m\alpha(\alpha Z)^{10}$ \, r.u.&$2.5\cdot 10^{-6}\, {\rm s}^{-1}$\\
${\rm 2p}\rightarrow {\rm 1s}$ & (E1M1) & $2.911\cdot 10^{-5}\, m\alpha^2(\alpha Z)^8$ r.u.&$9.68\cdot 10^{-6}\, {\rm s}^{-1}$\\
${\rm 2p}\rightarrow {\rm 1s}$& (E1E2) & $1.989\cdot 10^{-5}\,
m\alpha^2(\alpha Z)^8$ r.u.&$6.612\cdot 10^{-6}\, {\rm s}^{-1}$\\
\hline
\end{tabular}
\end{table}

\section{Conlusions}

In this paper analytical results for 2s, 2p levels decays for the
hydrogen-like atomic systems with one-, two- and three-photon
emission have been presented. All calculations were performed
within Pauli approximation utilizing the Coulomb Green function.
The emission processes were also calculated in the presence of an
external electric field. For the two-photon decays in the absence
of an external electric field the obtained results are in good
agreement with those of other calculations.

The two-photon decays E1E2 and E1M1 were evaluated with different
sets of quantum numbers (representations) for the emitted photon,
namely, parity and momentum or polarization vector ${\bf e}$ and
wave vector ${\bf k}$. Moreover, we have employed different forms
in combination with different gauges. The results do not differ in
magnitude by more then $0.1\%$ from fully relativistic values
which were obtained earlier. Recently a paper \cite{Amaro} did arrive where, in particular, 2p-1s E1M1 and E1E2 transition rates were evaluated for the H-like ions in the wide range of $Z$ values. For $Z=1$ the results of \cite{Amaro} are in agreement with our ones.

We have evaluated also the two-photon decay probabilities E1E2 and
E1M1 with the set of quantum numbers ${\bf e}$, ${\bf k}$ to
investigate the probability dependence on directions of the
photons emission. It allowed us to obtain the two-photon emission
processes in the presence of the external electric field. In
particular, we have demonstrated that interference terms in the
E1E1 and E1E2, E1M1 transitions appear, which depend linearly on
the external electric field.

The important result of our calculations is the prediction of a
characteristic difference in transition probabilities (spectra)
between H  and $\overline{{\rm H}}$ atoms in the presence an
external electric field, caused by the terms, linear in the
electric field. This effect was not yet discussed in literature.
The observation of this effect would also allow for drawbacks on
our understanding of fundamental symmetries in nature, i.e. the
CPT-symmetry: Any deviation from this result would provide a hint
for CPT violation from a low-energy physics scenario, i.e. atoms
in an external electric field.

Finally, we have compared the radiative correction evaluated in
\cite{Jentschura} with the electric field correction and
determined the magnitude of the electric field strength, when both
corrections become of the same order.

\begin{center}
Acknowledgements
\end{center}

The work of D.~S. and V.~S. was supported by the Non-profit
Foundation “Dynasty” (Moscow). D.~S., L.~L. and V.~S. also
acknowledge the support by RFBR grant grant Nr. 08-02-00026. V.~S.
acknoledges the support of St. Petersburg government. The authors
acknowledge finacial support provided by DFG, BMBF and GSI.  L. L. and D. S. acknowledge also the support by the Program of development of scientific potential of High School, Ministry of Education and Science of Russian Federation, grant $\aleph$2.1.1/1136.

\end{document}